\documentclass[12pt]{JHEP3}

\usepackage{amsmath,amssymb,amsfonts,amsthm}
\usepackage{bm,bbm}
\usepackage{graphicx}
\usepackage{esint}
\usepackage{color}

\newcommand{\be}{\begin{equation}}
\newcommand{\ee}{\end{equation}}
\newcommand{\ba}{\begin{eqnarray}}
\newcommand{\ea}{\end{eqnarray}}
\def\bs{\begin{subequations}}
\def\es{\end{subequations}}
\def\a{\alpha}
\def\b{\beta}

\def\cK{{\cal K}}

\def\cN{{\cal N}}
\def\cO{{\cal O}}

\def\cR{{\cal R}}

\def\p{\partial}

\newcommand{\m}{\mu}
\newcommand{\n}{\nu}

\newcommand{\half}{\tfrac{1}{2}}

\newcommand{\va}{\varphi}

\newcommand{\ct}{\tilde{c}}

\usepackage{MyJHEPStyle}
\newcommand{\gb}{\bar{g}}

\title{Fixed-Functionals of three-dimensional Quantum Einstein Gravity}

\author{Maximilian Demmel\\
PRISMA Cluster of Excellence \& Institute of Physics (THEP), \\
University of Mainz, Staudingerweg 7, D-55099 Mainz, Germany \\
E-mail: \email{demmel@thep.physik.uni-mainz.de}}

\author{Frank Saueressig\\
PRISMA Cluster of Excellence \& Institute of Physics (THEP), \\
University of Mainz, Staudingerweg 7, D-55099 Mainz, Germany \\
E-mail: \email{saueressig@thep.physik.uni-mainz.de}}

\author{Omar Zanusso\\
PRISMA Cluster of Excellence \& Institute of Physics (THEP), \\
University of Mainz, Staudingerweg 7, D-55099 Mainz, Germany \\
E-mail: \email{zanusso@thep.physik.uni-mainz.de}}

\abstract{We study the non-perturbative renormalization group flow of $f(R)$-gravity in three-dimensional Asymptotically Safe Quantum Einstein Gravity. Within the conformally reduced approximation, we derive an exact partial differential equation governing the RG-scale dependence of the function $f(R)$. This equation is shown to possess two isolated and one continuous one-parameter family of scale-independent, regular solutions which constitute the natural generalization of RG fixed points to the realm of infinite-dimensional theory spaces. All solutions are bounded from below and give rise to positive definite kinetic terms. Moreover, they admit either one or two UV-relevant deformations, indicating that the corresponding UV-critical hypersurfaces remain finite dimensional despite the inclusion of an infinite number of coupling constants. The impact of our findings on the gravitational Asymptotic Safety program and its connection to new massive gravity is briefly discussed.}

\preprint{MZ-TH/12-34}
\keywords{Quantum Gravity, Asymptotic Safety, Functional Renormalization Group}

\begin{document}


\section{Introduction}
Two central requirements any viable quantum theory of gravity has to
meet are
\begin{enumerate}
\item consistency and predictivity at high energies and
\item the emergence of a classical limit, where 
the phenomenological successes of classical general relativity are recovered.
\end{enumerate}
Weinberg's Asymptotic Safety scenario \cite{Weinberg:1980gg,Weinproc1} proposes that
these requirements can be satisfied within the realm of non-perturbative
quantum field theory. In this case, the UV-completion of gravity
is provided by a non-Gaussian fixed point (NGFP) of the renormalization
group (RG) flow which renders the theory safe from
unphysical divergences. Provided that the fixed point comes with a finite number of
UV-attractive (relevant) directions, this construction is as predictive as a 
``standard'', perturbatively renormalizable quantum field theory. 
Following the RG-flow emanating from the NGFP towards 
the IR, the second requirement can be fulfilled
 when the RG-trajectories enter into a regime
 where classical general relativity constitutes a good approximation. 
These ingredients are the foundation
of  ``Quantum Einstein Gravity'' (QEG), see \cite{Niedermaier:2006wt,Reuter:2007rv,robrev,Litim:2008tt,Reuter:2012id} for reviews.

The main technical tool for investigating Asymptotic Safety is the
functional renormalization group equation (FRGE) initially
constructed for scalar field theory \cite{Wetterich:1992yh} and subsequently generalized to gravity \cite{Reuter:1996cp}
\be\label{FRGE}
\frac{\diff }{\diff t} \Gamma_k[\Phi, \bar{\Phi}] = \half {\rm STr} \left[ \left( \frac{\delta^2 \Gamma_k}{\delta \Phi^A \delta \Phi^B} + \cR_k \right)^{-1} \frac{\diff}{\diff t} \cR_k \right] \, .
\ee
The FRGE constitutes an exact flow equation governing the change of the effective average action (EAA) $\Gamma_k$ upon integrating out a small shell of quantum fluctuations
of the field $\Phi$
centered at the IR-cutoff scale $k = \exp(t)$. Its solutions determine the flow of $\Gamma_k$ on the space of all action functionals.
Formally, they interpolate between a bare action at the UV-scale $k = \Lambda$ and the (standard) effective action (EA) $\Gamma = \Gamma_{k=0}$. 
Notably, applications of \eqref{FRGE}, often known as the Wetterich equation, are not limited to gravity, but also play an important role when exploring
non-perturbative physics in scalar field theory \cite{Bagnuls:2000ae,Berges:2000ew,Delamotte:2007pf,Rosten:2010vm}
and gauge theories \cite{Pawlowski:2005xe,Gies:2006wv,Braun:2011pp}.

The key strength of \eqref{FRGE} is that it allows the construction of non-perturbative approximations of the full RG-flow, meaning that they do not involve the expansion in a small coupling constant.
A non-perturbative approximation scheme, which is adopted quite frequently, is the truncation
of the full $\Gamma_k$ to a finite number of interaction monomials $\cO_i$ compatible with the symmetries of the theory
\be\label{ansatzGk}
\Gamma_k[\Phi, \bar{\Phi}] \simeq \int \diff^dx \sqrt{g} \sum_{i=0}^N \, u_i(k) \, \cO_i \, .
\ee
In the case of gravity, the $\cO_i$ are often taken as powers of the Ricci scalar $R$, and the truncation order has systematically been increased from 
the Einstein-Hilbert action ($N=1$) \cite{Reuter:1996cp,Dou:1997fg,Souma:1999at,Lauscher:2001ya,Reuter:2001ag,Litim:2003vp},
to $N=2$ \cite{Lauscher:2001rz,Lauscher:2002sq,Rechenberger:2012pm}, $N=6$ \cite{Codello:2007bd,Machado:2007ea}, and
$N=8$ \cite{Codello:2008vh}, up to $N=10$ \cite{Bonanno:2010bt}.
In \cite{Benedetti:2009rx}, the operator $\cO_3 = R_{\m\n\a\b} R^{\m\n\a\b}$
has been added to the $N=2$ case and non-local interactions have been considered in \cite{Reuter:2002kd,Machado:2007ea,Codello:2010mj,Satz:2010uu}.
Moreover, the study of quantum effects in the ghost-sector has been initiated in \cite{Eichhorn:2009ah}, a first investigation of 
 ``bimetric'' flows which discriminate between the background and fluctuation fields have been carried out in \cite{Manrique:2010am},
and RG-flows on surface-terms have been considered in \cite{Becker:2012js}.
Building on off-diagonal heat-kernel techniques, a general strategy for solving \eqref{FRGE} systematically order by order in the derivative expansion was proposed in \cite{Benedetti:2010nr,Saueressig:2011vn} and a physical explanation for Asymptotic Safety based on paramagnetic dominance 
 has been put forward in \cite{Nink:2012vd}. As a key common result, all these quite sophisticated computations found a suitable NGFP whose associated number of relevant directions could be as low as three \cite{Reuter:2012id}. Following a similar strategy, evidence for a non-trivial fixed point governing the IR-behavior of QEG have 
been reported in \cite{Donkin:2012ud,Nagy:2012rn,Litim:2012vz,Rechenberger:2012pm}.

On the technical side, most of these works include a finite number of couplings $\cO_i$.
In this case, one obtains the $N+1$ beta-functions $\beta_i$, capturing the scale-dependence of the $u_i$,
by substituting the ansatz \eqref{ansatzGk} into the FRGE and projecting the flow on the truncation subspace.
The fixed points of this finite-dimensional system are given by the zeros of the beta-functions and are thus found by solving the algebraic problem $\beta_i(u_i) = 0$.
The UV-relevant deformations of the fixed points are encoded in the stability matrix $B_{ij} \equiv \p_j \beta_i|_{\rm FP}$, obtained by linearizing the RG-flow at a given fixed point.
In terms of the stability coefficients $\theta_i$, defined as minus the eigenvalues of $B_{ij}$, the relevant deformations correspond to directions with positive $\theta_i$.

The next milestone in studying approximate solutions of \eqref{FRGE} consists in studying the RG-flow of $k$-dependent functions (fRG). 
In the language of \eqref{ansatzGk} this corresponds to including an infinite number of couplings.
At this level, the fixed points of the finite-dimensional truncations
 are promoted to fixed functions which arise as $k$-independent, regular solutions of partial differential equation (PDE) encoding the scale-dependence of the functions included in $\Gamma_k$. Relevant 
deformations can then be found by studying linear deformations of this PDE around the fixed function. Demanding regularity
of the deformation leads to a discrete set of relevant operators.  In scalar field theory this strategy 
has been first implemented in \cite{Wetterich:1992yh,Morris:1994ie} by deriving a non-linear PDE for the scale-dependence of the potential $V_k(\phi)$.
In this case the PDE includes up to two derivatives of $V$ with respect to $\phi$.
 Later on these ideas have been refined systematically \cite{Morris:1998da,Canet:2003qd,Litim:2010tt,Flore:2012ma}. 
 In \cite{Codello:2012sc} the local polynomial approximation of the scalar effective average action 
was used to analyze the dimension-dependent existence of universality classes.

The natural gravitational analogue of the local potential approximation in the scalar case is to approximate the gravitational part of the effective average action by a generic function of the 
Ricci scalar\footnote{The one-loop effective action for $f(R)$-gravity has been constructed in \cite{Cognola:2005de}.} 
\be\label{fRans}
\Gamma_k[g] = \int \diff^dx \sqrt{g} f_k(R) + \ldots \, .
\ee
A crucial difference between the scalar and gravitational theory is that diffeomorphism invariance does not allow to write down a potential for the metric fluctuations,
i.e., $f_k(R)$ doubles as both the kinetic and the potential term of the theory. Substituting \eqref{fRans} into \eqref{FRGE} the flow equation for $f_k(R)$ can be obtained by evaluating 
the FRGE on a spherical background, where all possible diffeomorphism invariant interaction monomials can be constructed from the Ricci scalar alone. A PDE  governing the $k$-dependence of $f_k(R)$ has been derived in \cite{Codello:2007bd,Machado:2007ea}. In contrast to the local potential approximation in the scalar case, this PDE
actually contains third derivatives of $f(R)$ with respect to $R$.
Using a different cutoff scheme, the asymptotic properties 
of the fixed functions arising in $f(R)$-gravity have recently been studied in \cite{Benedetti:2012dx}.\footnote{As will be explained in Sect.\ \ref{sect.2} the key technical difference between \cite{Benedetti:2012dx} and the present work is that we do not rely on a summation of the eigenvalues of the Laplacian on a sphere, but use the resummed form of heat-kernel \cite{Avramidi:2000bm}.}
The complex non-linear nature of the (ordinary) differential equation underlying the fixed functions made it very hard to find explicit solutions which are regular everywhere, however.

In this light, this paper studies a somewhat simpler system, the flow of $f_k(R)$ in conformally reduced gravity in $d=3$ spacetime dimensions. 
In the context of the FRGE, conformally reduced flow equations have been studied previously in \cite{Reuter:2008wj,Reuter:2008qx,Reuter:2009kq,Machado:2009ph,Bonanno:2012dg}. Here it was established that they lead to results quite similar to the full case, where the contributions of all metric fluctuations are included.\footnote{We expect that the conformal approximation in $d=3$ works even better than in $d=4$ dimensions, since $f(R)$-gravity
in the former case contains a single on-shell scalar degree of freedom, which is readily captured by this approximation.} Moreover, working in $d=3$ considerably simplifies the spectral sum 
of the Laplacians, so that traces of the form ${\rm Tr}[ e^{-s\Delta}]$ can be evaluated exactly. As it turns out, the 
``toy model'' emerging from this setting has all the key features of the PDE 
governing the scale-dependence of $f_k(R)$ in full QEG (third order derivatives, non-linearity, fixed and moving singularities) while 
still being simple enough that the combination of analytic and numerical methods allows the explicit construction of the corresponding 
fixed functions and their relevant deformations. In this sense, our analysis may be seen as an illustrative showcase, 
providing a comprehensive overview of mechanisms and structures expected in the (physically more interesting) case of
$f(R)$-gravity in four dimensions.

The remainder of our work is organized as follows. The non-linear partial differential equation governing the scale-dependence of $f_k(R)$ in three-dimensional conformally reduced QEG is derived in Sect.\ \ref{sect.2}. The scale-independent, regular solutions of this flow equation, which constitute the generalization of fixed points to infinite-dimensional truncations, are constructed in Sect.\ \ref{sect.3}. Their properties, including the number of UV-relevant deformations, are investigated in Sect.\ \ref{sect.4} and we discuss our findings in Sect.\ \ref{sect.5}.

\section{The FRGE on the three-sphere}
\label{sect.2}
We begin the section introducing the basic notions of how to implement a
conformally reduced gravitational path-integral.
The construction will be subsequently adapted to the formalism of functional renormalization group.
Finally, the flow equation of conformally reduced gravity on the three-sphere will be derived.

\subsection{Conformally reduced quantum gravity}

We start our investigation by discussing how conformally reduced gravity can be implemented around a definite background geometry.
The classical action $S_{\mathrm{grav}}[\gamma_{\mu\nu}]$ for pure gravity is in general a functional of the metric field $\gamma_{\mu\nu}(x)$.
Our aim is to construct an effective action where the quantum fluctuations of the metric are integrated out.
To achieve this in a covariant way, we use the background field method that allows to maintain background covariance of the path integral.
The basic idea of the background field formalism is to split the quantum metric $\gamma$ into a fixed, but otherwise arbitrary, background field $\bar{g}$ and a fluctuation field $h$
\eq{\label{split}
\gamma_{\mu\nu} = \bar{g}_{\mu\nu} + h_{\mu\nu}\,.
}
It is important to stress that the fluctuation field $h_{\mu\nu}$ is not an infinitesimal perturbation and hence it needs not to be ``small''.
The fluctuation field will be used as the integration variable of the functional integral that defines the quantum theory.

The theory is said to be ``conformally reduced'' if we simplify our model by restricting the quantum fluctuations to the conformal mode $\varphi = \gb^{\mu\nu}h_{\mu\nu}$.
In this case
\eq{\label{reduction}
h_{\mu\nu} \simeq \frac{1}{d}\gb_{\mu\nu}\varphi\,.
}
This is closely related to \cite{Hooft:2010nc}, where it is suggested that the integration of the conformal mode
should be performed first in a gravitational path integral.
An alternative procedure for the quantization of the conformal mode involves parametrizing the metric with a conformal factor over a fiducial background
\eq{
\gamma_{\mu\nu} = \chi^{2\nu}\gb_{\mu\nu}
}
for a certain exponent $\nu$.
The fluctuationfield $\chi$ is related to $\varphi$ via the identification $\chi^{2\nu} = 1+\frac{1}{d}\varphi$.
It has been shown that it is possible to treat $\chi$ as the quantum field, in fact
the quantum theory resulting from this kind of reduction is investigated in e.g. \cite{Reuter:2009kq}.
At the quantum level, however, the theories using $\chi$ and $\varphi$ as quantum fields are not necessarily equivalent as it was observed in \cite{Machado:2009ph}.
We will use $\varphi$ as quantum field, since it appears as gauge-invariant scalar mode in the standard York-decomposition of the metric.

The effective action is defined in a well-known way
via an integro-differential equation\footnote{Here we use the short hand notation $\int_x \equiv \int\diff^d x\; \sqrt{\gb}$.}
\eq{
 \label{eq:pathintegral}
 \mathrm{e}^{-\Gamma\left[ \phi;\gb\right]} = \int\Diff\varphi\;\E^{-S_{\mathrm{grav}} \left[\gb_{\mu\nu} + 
 \frac{1}{d} \gb_{\mu\nu}\varphi\right]+ \int_x  \frac{\delta \Gamma }{\delta \phi} \left( \varphi - \phi \right) },
}
where $\phi=\braket{\varphi}_J
$ is the expectation value of $\varphi$ in presence of the scalar source $J=\frac{\delta \Gamma }{\delta \phi}$.
We do not need to include ghosts or gauge fixing terms,
because diffeomorphism invariance is fully broken due to the fixing of the background metric $\bar{g}$.
In fact, the second order variation of any diffeomorphism invariant action,
like for example $S_{\rm grav}$, w.r.t.\ $\varphi$ generally does not admit any non-trivial zero modes. 

We analyze the RG behavior of the model by constructing a suitable scale dependent effective average action (EAA).
The EAA has a built-in scale dependence and can be obtained along the usual lines \cite{Reuter:1996cp,Machado:2007ea}.
First we add to the classical action in \eqref{eq:pathintegral} an infrared (IR) cutoff action, that is quadratic in the dynamical field and background-covariant
\eq{\label{cutoffaction}
\Delta S_k\left[ \phi;\gb \right] = \frac{1}{2} \int_x \phi(x) \mathcal{R}_k\left[\Delta ; \gb \right]\phi(x)\,.
}
Here $\Delta\equiv -\bar{D}^2$ is the Laplacian operator constructed with the background metric, so that background-covariance is always manifest.
Additionally, the cutoff operator is required to fullfil the boundary-conditions $\mathcal{R}_{k}\rightarrow 0$ for $k\rightarrow 0$ as well as  $\mathcal{R}_{k}\rightarrow k^2$ for $k\rightarrow \infty$.
This ensures that the EAA interpolates between a ``classical'' action at very high scales
and the EA $\Gamma\left[ \phi;\gb\right]$ defined in \eqref{eq:pathintegral} at $k=0$.

Infrared and ultraviolet modes are defined w.r.t.\ the operator $\Delta$:
the role of the cutoff term is to effectively suppress the propagation of IR modes, so that UV modes are progressively integrated out as the cutoff-scale $k$ decreases.
The EAA can then be defined through the modified path-integral
\eq{
 \label{eq:modifiedpathintegral}
 \mathrm{e}^{-\Gamma_k\left[ \phi;\gb\right]} = \int  \Diff\varphi\;   \E^{-S_{\mathrm{grav}} \left[\gb_{\mu\nu} +
 \frac{1}{d} \gb_{\mu\nu}\varphi\right]+ \int_x  \frac{\delta \Gamma_k }{\delta \phi} \left( \varphi - \phi \right) - \Delta S_k\left[ \varphi-\phi;\gb \right]}.
}
By construction, the EAA has a built-in scale-dependence $k$ and is therefore denoted $\Gamma_k[\phi,\bar{g}]$ with the additional label $k$.
If we differentiate the path integral \eqref{eq:modifiedpathintegral} with respect to RG time $t\equiv \ln k/k_0$, where $k_0$ is an arbitrary reference scale, we obtain the flow equation
\eq{\label{eq:floweq}
\frac{\diff}{\diff t} \Gamma_k[\phi;\bar{g}] = \frac{1}{2} \Tr{\left(\Gamma_k^{(2,0)}\left[\phi;\bar{g} \right] + \mathcal{R}_k\right)^{-1}\frac{\diff}{\diff t}\mathcal{R}_k},
}
which is this system's specific form of \eqref{FRGE}.
We used the short hand $\Gamma_k^{(2,0)}\left[\phi;\bar{g} \right]=\frac{\delta^2 \Gamma_k}{\delta \phi \delta\phi}$, to indicate derivatives with respect to the first argument.

%
%
%
%
\subsection{Computing the variations}
%
%
%
%
%
%
Having the flow equation at our disposal, we now make an ansatz for the EAA.
For the purpose of the present work, we resort to the single-metric approximation, i.e. 
\eq{\label{backgroundapproximation}
\Gamma_k[\phi;\bar{g}] = \Gamma^{\mathrm{grav}}_k[g]\,,
}
with $\braket{\gamma}= g = \gb + \frac{1}{d}\gb\phi$ being the expectation value of the full quantum metric.
For this EAA we now assume a generic $f(R)$ ansatz
\eq{\label{eq:ansatz}
\Gamma^{\mathrm{grav}}_k[g] = \int\mathrm{d}^d x\sqrt{g}\; f_k(R)\,,
}
where $R$ is the scalar curvature of the metric $g$.
The usual prefactor $(16\pi G_k)^{-1}$, involving the (scale-dependent) dimensionful Newton's constant $G_k$, is absorbed into the now $k$-dependent function $f_k(R)$.
Since the second functional derivative of $\Gamma_k$ is needed to compute the flow equation \eqref{eq:floweq},
we expand the EAA in a series in $h$ around the background field
\eq{\label{eq:seriesexpansion}
\Gamma_k[\bar{g} + h] = \Gamma_k[\bar{g},\bar{g}] + \mathcal{O}(h) + \Gamma_k^{\mathrm{quad}}[h,\bar{g}]+ \mathcal{O}(h^3)\,.
}
The conformal reduction is achieved by setting the fluctuations to be $h_{\mu\nu} = \frac{1}{d} \bar{g}_{\mu\nu} \phi$.
Performing the second order expansion of \eqref{eq:ansatz} gives
\eq{
\delta^2 \Gamma_k[g] = \int \diff^dx \left[ f_k(R)\, \delta^2\sqrt{g} +f'_k(R)\,(2(\delta\sqrt{g})(\delta R) + \sqrt{g}\delta^2 R) + \sqrt{g}f''_k(R)\,(\delta R)^2 \right],
}
where the prime denotes a derivation with respect to the scalar curvature, i.e. $f'_k(R) = \frac{\del f_k(R)}{\del R}$.
The computation is simplified further by choosing a spherically symmetric background geometry which e.g. implies $\bar{D}_\mu f_k(\bar{R})=0$.
Computing the second order expansion in the conformal mode $\phi$ the quadratic part of the series \eqref{eq:seriesexpansion} becomes
\eq{\label{eq:gammaquad}
\Gamma^{\rm{quad}}_k[\phi,\bar{g}] = \int\mathrm{d}^d x\sqrt{\bar{g}}\; \frac{1}{2} \phi(x) \mathcal{K}_k \phi(x)\,,
}
with kernel
\be\label{G2var}
\begin{split}
 \mathcal{K}_k
 =\Gamma^{(2,0)}_k[0,\bar{g}]
 = \frac{1}{4d^2}&\left[ 4(d-1)^2 f''_k \bar{D}^4 - 2(d-1)\left((d-2)f'_k - 4 \bar{R} f''_k \right)\bar{D}^2 \right. \\ & \, + \left. (d-2)(f_k d - 4 \bar{R} f'_k) + 4 \bar{R}^2 f''_k  \right]\,,
\end{split}
\ee
where $\bar{R}$ is the scalar curvature computed from the background metric $\bar{g}$.

Now it is time to construct the cutoff operator $\mathcal{R}_k$ in such a way that the propagation of the conformal mode $\phi$ is IR-suppressed.
The Hessian \eqref{G2var} is a function of the covariant Laplacian $\Delta = -\bar{D}^2$ of the form
\eq{
\Gamma^{(2,0)}_k = f(-\bar{D}^2;\dots) \,,
}
where all other dependences are denoted by dots for brevity.
We choose the cutoff implicitly such that the sum of the kernel $\mathcal{K}$ and the cutoff operator satisfies
\eq{
\Gamma^{(2,0)}_k + \mathcal{R}_k = f(-\bar{D}^2 + R_k;\dots) \,.
}
Thus $\mathcal{R}_k$ maintains the same tensorial structure as $\Gamma^{(2,0)}_k$, but dresses every Laplacian according to
\eq{
\Delta \rightarrow P_k\equiv \Delta + R_k \,.
}
The function $R_k=R_k(\Delta)$ is the profile function for the cutoff that contains the details of the IR-modes suppression and will be specified later.
The function $P_k=P_k(\Delta)$ plays the role of a IR-modified propagator for the scalar modes.
The cutoff action $\Delta_k S_{\rm grav}$
are determined explicitly by these requirements
\eq{\label{eq:explicitcutoff}
 \begin{split}
 \Delta_k S_\mathrm{grav}
 = \frac{1}{2} \int\mathrm{d}^d x \sqrt{\bar{g}}\; \frac{1}{4d^2}\phi(x) \left[4(d-1)^2f''_k \left(P_k^2 - \bar{D}^4 \right) \right. \\ +\left. 2(d-1)\left( (d-2)f'_k - 4 \bar{R} f''_k \right)R_k \right]\phi(x)\,.
 \end{split}
}
Using \eqref{G2var} and \eqref{eq:explicitcutoff} to compute the flow equation \eqref{eq:floweq} in the limit $\phi=0$ yields
\eq{\label{eq:finalfloweq}
\frac{\diff}{\diff t} \Gamma_k[0, \bar{g}] =  \frac{1}{2} \Tr{W\( \Delta\)} ,
}
with $\Tr{W}= \Tr{\mathcal{N}\mathcal{A}^{-1}}$ and
\begin{subequations}
 \label{eq:NoverA}
  \eq{
  \mathcal{N} =&\frac{\diff}{\diff t} \left[ \left( 4(d-1)^2 f''_k \left(P^2_k - \bar{D}^4 \right)  + 2(d-1) \left( (d-2)f'_k - 4 \bar{R} f''_k \right) R_k\right) \right],\\
 \begin{split}
  \mathcal{A} =&
  \left[ 4(d-1)^2 f''_k P_k^2 + 2(d-1)\left((d-2)f'_k - 4 \bar{R} f''_k \right)P_k\right. \\
  & \left. + (d-2)(f_k d - 4 \bar{R} f'_k) + 4 \bar{R}^2 f''_k  \right]\,.
 \end{split}
}
\end{subequations}
It is easy to show that \eqref{eq:finalfloweq} coincides with the flow equation for $f(R)$-gravity derived in \cite{Machado:2007ea}
in the limit in which the physical scalar is the only dynamical mode of the gravitational fluctuations.

In three spacetime dimensions $(d=3)$ the function under the trace reads simply
\eq{\label{eq:finalfloweq3sphere}
W(z)=\frac{\dfrac{\diff}{\diff t}\left(g_k\left(P_k^2 - z^2 \right) + \tilde{g}_k R_k \right)}{g_k P^2_k + \tilde{g}_k P_k +w_k},
}
with coefficients
\eq{\begin{split}\label{coeffs}
g_k &= 16 f''_k,\\
\tilde{g}_k &= 4 f'_k - 16f''_k \bar{R},\\
w_k &= 4 \bar{R}^2f''_k - 4 \bar{R} f'_k +3f_k\,.
\end{split}
}
%
%
%
%
%
\subsection{Evaluating the operator traces}
%
%
%
The final step to evaluate the flow equation for the function $f_k(R)$ is to perform
the functional trace of the function $W$ of the operator $\Delta$ appearing the r.h.s.\ of \eqref{eq:finalfloweq}.
There are in principle different strategies to tackle this task. We briefly outline two of them that give different results, but only because they ultimately correspond to different definitions of how a functional trace should be computed.

The straightforward possibility is to compute the trace directly by defining it as the spectral sum
\eq{\label{eq:benedettiway}
\Tr{W(\Delta)} = \sum_n W\(\lambda_n\),
}
where $\left\{ \lambda_n \right\}$ are the eigenvalues of the Laplace operator $\Delta$ on the sphere.
This has been done in \cite{Benedetti:2012dx} applied to the case of four spacetime dimensions (four-sphere).
However this procedure has its limitations. In particular, if $W$ is a distribution (instead of a smooth function),
the final result for the trace is a distribution as well and therefore it may lead to a flow
equation that has some behavior that is hard to interpret from a physical point of view
and may require further manipulation \cite{Benedetti:2012dx}.

We decided, instead, to define the trace of any function (and distribution)
of the Laplacian through the heat-kernel operator $\E^{-s\Delta}$.
The trace of a general function of $\Delta$ is formally related to the heat-kernel
\eq{\label{eq:ourway}
\Tr{W(\Delta)} = \int\limits_0^\infty\diff s\;\tilde{W}(s)\Tr{ \mathrm{e}^{-s\Delta} }
}
via the inverse Laplace transform of the function itself defined by
\eq{
\tilde{W}(s) = \mathcal{L}^{-1}\left[ W \right](s)\,.
}
In contrast to the definition \eqref{eq:benedettiway},
it is important to realize that there is no difficulty in principle to use the heat-kernel definition of the trace when $W$ is a distribution. In fact, any distribution can be seen as the limit of a sequence of functions
and therefore it is understood that the inverse Laplace transform appearing in \eqref{eq:ourway}
is the limit of the inverse Laplace transforms of the elements of the sequence.
It will be evident in the following that, in our applications, this limit tends to a well-defined smooth function.

The heat-kernel operator $\E^{-s\Delta}$ occurring in the trace is a well-known object in both mathematical and physical literature.
Various techniques have been developed to compute it, of which we cite only a few \cite{Vilkovisky:1992za,Avramidi:2000bm,Groh:2011dw,Codello:2012kq}.
A standard approach, aiming at computing the trace, is to start with an asymptotic expansion for the matrix elements of the diagonal heat-kernel
\eq{\label{eq:earlytime}
K(s;x,x) \equiv \braket{x|\E^{-s\Delta}|x} = \frac{1}{\(4\pi s\)^{d/2}}\sum_n a_n(x) s^n,
}
where the $a_n$ are known as deWitt coefficients.
If we restrict our attention to a scalar theory on a sphere (Euclidean de Sitter space),
the asymptotic series \eqref{eq:earlytime} can be resummed up to non-analytic terms.
This is achieved by either using a covariant expansion of the deWitt coefficients and performing the resummation \cite{Avramidi:2000bm},
or by means of asymptotic expansions of the Green function at coinciding points \cite{Dowker:1975tf}
from which one can read off the expansion of the heat-kernel itself \cite{Avramidi:2000bm}.
In $d=3$ the exact resummation of \eqref{eq:earlytime} is particularly simple \cite{Avramidi:2000bm}
\eq{\label{eq:heat-kernel}
K(s;x,x)= \frac{1}{(4\pi s)^\frac{3}{2}} \E^{  \frac{1}{6}R s}\,.
}
We notice that the result \eqref{eq:heat-kernel} can also be obtained performing the spectral sum \eqref{eq:benedettiway} for the case $W(z)=\E^{-sz}$
using the Euler-MacLaurin formula for the summation of a series\footnote{We are grateful to R. Percacci for pointing this out.}.
This observation implies that the summation does not commute with the operation of inverse Laplace transform, leading to a crucial difference between the definitions \eqref{eq:benedettiway} and \eqref{eq:ourway}.
Whether one method is ``better'' than the other is debatable since they are simply different definitions.
However, the regularity properties inherent in definition \eqref{eq:ourway} have the clear advantage of making the functional trace \eqref{eq:finalfloweq}
more regular as a function of the cutoff and therefore less scheme dependent.

Evaluating the trace in \eqref{eq:finalfloweq}, using the inverse Laplace transform together with the heat-kernel \eqref{eq:heat-kernel} gives
\eq{\label{eq:3dtrace}
\Tr{W(\Delta)} &=\int_0^\infty\diff s\; \tilde{W}(s) \Tr{\E^{-s\Delta}}   \nonumber\\
&= \frac{1}{(4\pi)^{\frac{3}{2}}} \int\diff^3x \sqrt{\gb}\; Q_{\frac{3}{2}}\left[\hat{W}\right]\,,
}
where $\hat{W}(x) \equiv W\left(x-\frac{R}{6}\right)$.
Moreover $Q_n$ denotes the general Mellin-transform of $W$ defined by
\eq{ \label{eq:mellin}
Q_n\left[ W \right] &\equiv \int_0^\infty\diff s\;s^{-n} \tilde{W}(s)
= \frac{1}{\Gamma(n)}\int_0^\infty\diff z\;z^{n-1}W(z)\,,
}
where the second equality is true only for $n\ge 0$.
The proof of \eqref{eq:3dtrace} uses the fact that $\mathrm{e}^{-s\alpha}$ represents the translation operator in the Laplace representation of the space of functions.
In fact, it is easy to prove that
\eq{
Q_n \left[ W(z+\alpha) \right] = \int_0^\infty\diff s\;s^{-n} \tilde{W}(s)\mathrm{e}^{-s\alpha}\,,
}
where the notation means that the Mellin-transform is performed with respect to the argument $z$, instead of the full argument of $W$.

In our particular case we use the definition \eqref{eq:mellin} for the case $n=\frac{3}{2}$ to evaluate
the Mellin-transform appearing in \eqref{eq:3dtrace} as
\eq{\label{qthreehalf}
Q_\frac{3}{2}\left[ \hat{W} \right]&= \frac{1}{\Gamma(\frac{3}{2})}\int_0^\infty\diff z\;z^{\frac{1}{2}}\,W\left(z -\frac{R}{6} \right).
}
We stress that up to now the derivation did not rely on any specific property of the profile function $R_k$.
%
%
%
\subsection{Cutoff and threshold functions}
%
%
%
%
%
%
%
Up to now, the profile function $R_k$ of the cutoff-operator was kept unspecified.
It is convenient to choose it to be the so-called optimized cutoff function, which is the distribution
\eq{\label{eq:optcutoff}
R_k(z) &= \( k^2 - z \) \theta\left( k^2 -z\right),\quad z = \Delta\,,
}
that was originally developed with the purpose of optimizing the convergence of the RG-flow \cite{Litim:2001up}.
Owed to its technical simplicity, this choice allows many explicit exact computations in the context of functional RG and, in particular, of integrals like \eqref{qthreehalf}.
These integrals are often called ``threshold functions'' since they posses a threshold-like structure in the form of a non-trivial denominator.
It is useful to compute the scale derivative of $R_k$ and the function $P_k$ already at this stage
\eq{\begin{split}
 & \frac{\diff}{\diff t} R_k(z) = 2k^2 \theta\left( k^2 -z\right),\\
 & P_k = z + R_k = k^2\theta\(k^2-z\) + z \,\theta\(z-k^2\).
\end{split}
}
Due to the presence of the theta functions, the domain of integration inside
\eqref{qthreehalf} reduces to $[0,k^2-c]$, with $c=-R/6$.
In the particular case of the three-sphere $R\geq 0$ and therefore the domain $[0,k^2-c]$ is always non-empty.
Using $W$ as defined in \eqref{eq:finalfloweq3sphere}, we obtain
\eq{\label{qthreehalf2}
Q_{\frac{3}{2}}\left[W(z+c)\right] =\frac{1}{\Gamma\( \frac{3}{2} \)} \int_0^{k^2 - c}\!\diff z\, z^{\frac{1}{2}}\, \frac{\tilde{u}_kz^2+\tilde{v}_k z + \tilde{w}_k}{g_k k^4 + \tilde{g}_k k^2 + w_k}\,,
}
with coefficients given in \eqref{coeffs} and
\eq{\begin{split}
 \tilde{u}_k &= -\dot{g}_k,\\
 \tilde{v}_k &= -2\dot{g}_k c -\dot{\tilde{g}}_k,\\
 \tilde{w}_k &= \dot{g}_k k^4 - \dot{g}_k c^2 + 4k^4 g_k + \dot{\tilde{g}}_k(k^2-c) + 2k^2\tilde{g}_k\,.
 \end{split}
}
For the optimized cutoff \eqref{eq:optcutoff} the $z$-integration can be carried out analytically.
We denote the basic integrals with a notation similar to \eqref{eq:mellin}, but take into account that the $z$-integration is now bounded to the upper limit $k^2-c$
\eq{\label{eq:finalQfunctional}
 \tilde{Q}_n\left[z^m\right] \equiv
 \frac{1}{\Gamma(n)}\int_0^{k^2-c}\,\diff z\, z^{n+m-1} = \frac{1}{\Gamma(n)}\frac{\left(k^2-c\right)^{n+m}}{n+m}\,.
}
Concretely, these integrals are needed in the cases $n=\frac{3}{2}$ and $m=0,1,2$.
%
%
\subsection{Flow equation}
%
%
%
%
At this stage we have all the ingredients to write down the full flow equation for the $f(R)$-truncation.
Inserting our ansatz \eqref{eq:ansatz} into the l.h.s.\ of \eqref{eq:finalfloweq}, one simply gets
\eq{\label{eq:finalfloweqLHS}
\dot{\Gamma}_k= \int\diff^3x\sqrt{\gb}\; \dot{f}_k\,,
}
where the dot will always denote the derivative w.r.t.\ the RG time $t=\log \left( k/k_0 \right) $.
Combing our previous results for the functional trace \eqref{qthreehalf2} and the definitions \eqref{eq:finalQfunctional}, the r.h.s.\ of \eqref{eq:finalfloweq} can be obtained
\eq{\label{eq:finalfloweqRHS}
\frac{1}{2} \Tr{W(\Delta)} &= \frac{1}{2} \frac{1}{(4\pi)^{\frac{3}{2}}} \int\diff^3x\sqrt{\gb}\;Q_{\frac{3}{2}}\left[W(x+c)\right]\nonumber\\
&=\frac{1}{2}\frac{1}{(4\pi)^{\frac{3}{2}}}  \int\diff^3x\sqrt{\gb}\;
\frac{\tilde{u}_k \tilde{Q}_\frac{3}{2}\left[z^2\right] + \tilde{v}_k \tilde{Q}_\frac{3}{2}\left[z\right] + \tilde{w}_k \tilde{Q}_\frac{3}{2}\left[1\right]}{ u_k k^4 + v_k k^2 + w_k}\,.
}
Equating \eqref{eq:finalfloweqLHS} and \eqref{eq:finalfloweqRHS}, making use of the explicit form of the integrals \eqref{eq:finalQfunctional}, yields the flow equation for $f_k$
\eq{\label{fdot}
\dot{f}_k=\frac{\left(6 k^2+R\right)^{3/2} \left( \alpha_1 f'_k + \alpha_2 f''_k + \alpha_3 \dot{f}'_k + \alpha_4 \dot{f}''_k  \right)}{5670 \sqrt{6} \pi ^2 \left(3 f_k+4 \left(\left(k^2-R\right) f'_k+\left(-2 k^2+R\right)^2 f''_k\right)\right)}
}
with
\eq{
\begin{split}\begin{aligned}
\alpha_1 &= 630 k^2 \, ,  &\alpha_2 =& 2520 \left(2 k^4-k^2 R\right) \, , \\
\alpha_3 &= 21 (6 k^2 + R) \, ,  &\alpha_4 =& 720 k^4 - 432 k^2 R - 92 R^2 \, .\\ 
\end{aligned}\end{split}
}
This is the desired partial differential equation governing the scale dependence of $f_k$.

For the purpose of analyzing the flow, we express $R$ and $f_k(R)$ in terms of dimensionless quantities.
Therefore we introduce the dimensionless analogs of $R$ and $f_k(R)$
\eq{\label{dimless1}
R \equiv k^2 r\,, &\qquad 
f_k(R)\equiv k^3 \varphi_k(R/k^2)\,.
}
The derivatives of $f_k$ and $\varphi_k$ are related by
\eq{\begin{split}\label{dimless2}\begin{aligned}
f'_k &= k \varphi'_k\,,\quad &  \dot{f}_k &=  k^3 \left(  \dot{\varphi}_k  + 3\varphi_k - 2r\varphi'_k \right)\,,\\
f''_k &= k^{-1}\varphi''_k\,,\quad  &  \dot{f}'_k &=  k \left(  \dot{\varphi}'_k  + \varphi'_k - 2r\varphi''_k \right)\,, \\
&&\dot{f}''_k &=  k^{-1} \left(  \dot{\varphi}''_k  - \varphi''_k - 2r\varphi'''_k \right)\,.
\end{aligned}\end{split}
}

We can now use \eqref{dimless1} and \eqref{dimless2} in \eqref{fdot} to write down the complete flow equation in terms of dimensionless quantities
\eq{\label{pdgl}
\dot{\varphi}_k + 3\varphi_k - 2r\varphi'_k = \frac{1}{\pi^2}\left( 1 + \frac{r}{6}  \right)^\frac{3}{2}\frac{c_1\varphi'_k +  c_2 \varphi''_k +  c_3\dot{\varphi}'_k + c_4\left(\dot{\varphi}''_k-2r\varphi'''_k\right)}{3\varphi_k+4(1-r)\varphi'_k + 4\left(2-r\right)^2\varphi''_k}\, ,
}
where the $c_i$ are polynomials of the dimensionless curvature $r$
\be\label{ccoeff}
\begin{split}
\begin{aligned}
c_1 &= \frac{1}{45} \left( 36 + r \right) \, , &  c_2 =& \frac{2}{189} \left(5 r^2 - 234 r + 432 \right) \, , \\
c_3 &= \frac{1}{45} \left( 6 + r \right) \, , &  c_4 =& \frac{4}{945} \left(6 + r \right) \, \left(30 - 23 r \right) \, . 
\end{aligned}
\end{split}
\ee
Eq.\ \eqref{pdgl} constitutes the central result of this section. It will be the starting point for our search for fixed functions of our functional RG system.
%
%
%
%
%
\section{Constructing the fixed functions of the RG-flow}
\label{sect.3}
%
%
%
%
%
We now carry out a systematic search for fixed functions of the gravitational
RG-flow, which arise as regular, scale-independent solutions of eq.\ \eqref{pdgl}.
These solutions generalize the fixed points, found on finite dimensional truncations of theory space, to the infinite-dimensional theory space of $f(R)$-gravity.
%
%
%
%
\subsection{General structure of the fixed function equation}
%
%
%
Imposing $k$-independence of the solution, eq.\ \eqref{pdgl} reduces 
to a non-linear ordinary differential equation of third order
\be\label{fpeq}
3 \va - 2 r \va' = \frac{1}{\pi^2} \left( 1 + \tfrac{r}{6} \right)^{3/2} \, \frac{c_1 \, \va' + c_2 \, \va''  - 2 r c_4  \va''' }{4  \, (2-r)^2 \, \va'' + 4 \, (1-r) \, \va' + 3 \va } \, ,
\ee
where, in a slight abuse of notation, $\va = \va(r)$. The non-linearity of this equation makes it very difficult to 
find its solutions analytically. Thus we resort to a combination
of analytic and numerical methods to construct its solutions.

In order that a solution constitutes a valid fixed function, we require that it is regular on the entire
positive real line $r \in [0, \infty[$. In terms of the dimensionful quantities, this condition reflects the requirement that the fixed function
exists for all values of $k$: keeping the dimensionful curvature $R$ of the background fixed, $k \in [0, \infty [\; \mapsto\; r =R/k^2 \in\ ]\infty, 0]$.
As we shall see in the following, this regularity condition puts severe constraints on the initial data characterizing
the solutions of \eqref{fpeq}.

We start our search by solving eq.\ \eqref{fpeq} for $\va'''$. Inserting $c_4$ from eq.\ \eqref{ccoeff},
we obtain
\be\label{FFsolved}
\va''' = \frac{945 \, \cN(\va, \va',\va'', r)}{8 \left(6 + r \right) \, \left(30 - 23 r \right) r} \, ,
\ee
where the numerator  
\be\label{Nfct}
\cN = \left(\cN^{\rm lin} - \pi^2 \left( 1+r/6 \right)^{-3/2} \cN^{\rm quad} \right)
\ee
 has been decomposed into parts linear and quadratic in $\va$ and its derivatives
\be\begin{split}
\cN^{\rm lin} \equiv & \, c_1 \va' + c_2 \va'' \, , \\
\cN^{\rm quad} \equiv &  \Big( 4  \, (2-r)^2 \, \va'' + 4 \, (1-r) \, \va' + 3 \va \Big) \Big( 3 \va - 2 r \va' \Big) \, . 
\end{split}
\ee

The solutions of \eqref{FFsolved} are characterized by three initial conditions for $\va, \va', \va''$ at an initial point $r_{\rm init}$. 
This is most easily seen by expressing the solution in a Taylor-series expansion around $r_{\rm init}$. Substituting this expansion
into \eqref{fpeq}, the higher-derivative terms in this expansion are fixed in terms of the three initial conditions entering the r.h.s.

Inspecting the r.h.s.\ of \eqref{FFsolved}, we observe the appearance of ``fixed singularities''. The denominator vanishes
at
\be\label{poles}
r = 0 \, , \qquad r = \infty \, , \qquad \mbox{and} \quad r = r_{\rm sing} \equiv \frac{30}{23} \approx 1.30 \, .
\ee  
Expanding the solution at these singular points shows that the poles at $r=0$ and $r_{\rm sing}$ are of first order
while the pole at infinity is second order.
These fixed singularities motivate to divide the positive real line into two regions
\be
\begin{split}
\cR_0: \;  0 \le r \le r_{\rm sing} \, , \qquad 
\cR_\infty: \; & r_{\rm sing} \le r < \infty \, ,
\end{split}
\ee
and use the coordinate $ x \equiv r^{-1}$ on $\cR_\infty$.
In addition to these fixed singularities, the non-linear nature of \eqref{fpeq}
can give rise to solutions, terminating in ``moving singularities''. In this case, the solution terminates at some $r = r_{\rm term} > 0$
by developing a logarithmic singularity
\be\label{logsing}
\va(r) \propto \log|r - r_{\rm term}| \, . 
\ee
This second type of singular behavior can easily be detected when constructing explicit solutions numerically.
Based on this discussion of the general structure, we now proceed by explicitly constructing the regular solutions of \eqref{fpeq} in the next subsections.

\subsection{The Gaussian fixed function}
We start by constructing the ``trivial'' or Gaussian fixed function, which generalizes the Gaussian fixed points (GFP) associated with the free or Gaussian theory
present on any theory space. This ``trivial'' solution of the fixed function equation is obtained as follows. We first introduce a coupling constant $c$, which controls the
interactions of the theory. At the level of the function $\va(r)$ this coupling is introduced via the rescaling 
\be\label{resc}
\va[r(g_{\m\n})] \rightarrow \tfrac{1}{c} \va[r(g_{\m\n})].
\ee
Subsequently, we chose a fixed background $\bar{g}_{\m\n}$ and express the metric $g_{\m\n}$ through the metric fluctuation $h_{\m\n}$ around this background 
\be
g_{\m\n} = \bar{g}_{\m\n} + \sqrt{c} \, h_{\m\n} \, . 
\ee
Substituting this expression into the rescaled $\va$ and expanding in $h_{\m\n}$ the order quadratic in $h_{\m\n}$, encoding the propagator of the theory, is independent of $c$.
Higher order terms, forming the interaction vertices, are proportional to a positive power of $\sqrt{c}$. Thus, in the limit $c \rightarrow 0$ the action becomes quadratic in $h_{\m\n}$, so that the corresponding path integral is Gaussian.     

When analyzing the limit $c \rightarrow 0$ of the fixed function equation \eqref{fpeq}, we exploit that its l.h.s.\ and r.h.s.\ are homogeneous of degree one and zero under the rescaling \eqref{resc}, respectively. Hence, in the limit of vanishing $c$ the r.h.s., which contains the trace-contribution of the Wetterich equation, decouples and one obtains the Gaussian fixed point equation
\be\label{GFPeq}
3 \va - 2 r \va' = 0 \, . 
\ee
This equation has the solution 
\be\label{GFP}
\va_{\rm GFP}(r) = a \, r^{3/2},
\ee  
where $a$ is a free integration constant. This constant just sets the overall normalization of the free propagator and has no physical significance.

Owed to the decoupling of the quantum corrections, the result \eqref{GFP} is easily generalized to space-times with dimension $d$.
The numerical coefficients in \eqref{GFPeq} just encode the canonical mass dimensions of the space-time integral and Ricci-scalar,
so that the general result can be obtained by substituting $3 \rightarrow d$ in the first term. Thus the $d$-dimensional Gaussian fixed point solution is given by
\be\label{GFPgen}
\va_{\rm GFP}(r) = a \, r^{d/2}.
\ee
For $d=4$, this general result reduces to the Gaussian fixed point solution found in \cite{Benedetti:2012dx}. 
Notably, the solutions \eqref{GFPgen} are distinguished by the property that the dimensionful coupling constants multiplying the
$R^{d/2}$-term are powercounting marginal, i.e., of mass-dimension zero. This is expected from the fact that, for the free theory at the Gaussian fixed point, there are
no quantum corrections that could lead to an anomalous dimension of the operators.

\subsection{Non-Gaussian fixed functions}
\label{sect:3.3}
The remainder of this section is devoted to
the search of non-trivial regular solutions
of eq.\ \eqref{fpeq}. The analysis of the singularity structure \eqref{poles} showed
 that the fixed function equation 
possesses a double pole at $r = \infty$, while the pole at $r=0$ is of first order.
We thus expect that imposing the regularity condition at
large $r$ will fix two of the three initial conditions characterizing the solutions,
while an analogous analysis at small $r$ may leave two parameters undetermined.
Hence we will start in the IR ($r \gg 1$) 
and continue our analysis towards the UV ($r \ll 1$).

\subsubsection*{Asymptotic solutions at large $r$}
We start with fixing the asymptotic behavior of the fixed functions for $r \rightarrow \infty$. For
this purpose, we make the ansatz
\be\label{asymexp}
\va \simeq c \, r^\alpha \, ( 1 + u_1 \, r^{-1} + \ldots ) \, , 
\ee
where the exponent $\alpha$ determines the asymptotic behavior of the solution, and  
the dots denote subleading terms.
Substituting this ansatz into \eqref{fpeq},
the asymptotic expansion yields
\be\label{asyexp}
c \, r^\alpha \, (3- 2 \alpha) = \left\{
\begin{array}{ll}
 \tfrac{\alpha \, (92 \alpha -113) }{5670 \sqrt{6} \pi^2 \, (2\alpha-1)} \, r^{3/2} + \ldots & \qquad \alpha > 3/2 \\[1.5ex]
 \tfrac{945 - 67 u1}{34020 \, \sqrt{6} \, \pi^2} \, r^{3/2} + \ldots & \qquad \alpha = 3/2 \, .
\end{array}
\right. 
\ee
When expanding the r.h.s.,
the limit $\alpha = 3/2$ is non-trivial, since in this case the leading terms in the numerator
and denominator both vanish, such that the coefficient $u_1$ appearing at subleading order actually enters into the equation.
Comparing powers of $r$, \eqref{asyexp} establishes that for $\alpha > 3/2$ the only solution of this equation is $c = 0$.
Thus the highest power of $r$ allowed in the asymptotic expansion is  $\alpha = 3/2$. This case allows the construction of a non-trivial solution ($c \not = 0$) 
by adjusting the subleading coefficient $u_1$. Notably, the asymptotic behavior $r^{3/2}$ reproduces the large $R$ limit of the one-loop 
effective action in three dimensions \cite{Avramidi:2000bm}.

Together with the square-root appearing in the fixed function equation, this asymptotic behavior motivates redefining
\be\label{varedef}
\va(r) = \left( 1 + r/6 \right)^{3/2} \, y(r) \, .
\ee
As it turns out, the prefactor in \eqref{varedef} captures the asymptotic behavior of $\va(r)$ for both
small and large values of $r$. The profile function $y(r)$ then interpolates continuously between 
these two asymptotic regimes.

In order to proceed with the analysis, we adapt the differential equation for the profile function $y(r)$ to the patch $\cR_\infty$ by changing variables
$x = 1/r$. The function $v(x) \equiv y(1/x)$ then satisfies the non-linear third order equation
\be\label{vdgl}
v''' = \frac{\tilde{\cN}(v,v',v'',x)}{240 \, x^2 \, (1+6 x)^4 \, ( x-x_{\rm sing})} \, .
\ee
Here the prime denotes a derivative with respect to the argument $x$, $x_{\rm sing} = 1/r_{\rm sing} = 23/30$,
and the numerator is given by
\be\label{Ntilde}
\begin{split}
\tilde{\cN} = & \, \ct_0 \, \left( 18 \, (19-558x) \, v - \ct_0 \, \ct_1 \, v' + 2 \, x \, \ct_0^2 \ct_2 \, v'' \right) \\
& \, - 3780 \pi^2 x \, \left(9 v + \ct_0 v' \right)\,  \left( (3 - 78 x) \, v - 2x \, \ct_0 \ct_3 \, v' - 2 x \, \ct_0^2 \, (1-2x)^2 \, v'' \right) \, .
\end{split}
\ee
The coefficients $\ct_i$ are polynomials in $x$
\be
\begin{array}{ll}
\ct_0 =  \, 1 + 6x \, , & 
\ct_1 =  \, 67 - 1434 \, x - 72360 \, x^2 + 103680 \, x^3 \, , \\
\ct_2 =  \, 113 + 4302 \, x - 6480 x^2 \, , \quad \quad& 
\ct_3 =  \, 21 - 58 x + 48 x^2 \, .
\end{array}
\ee

As already anticipated, the denominator of \eqref{vdgl} has a double zero at $x=0$.
Insisting that the asymptotic behavior of the solution is given by \eqref{varedef}, 
this divergence fixes two of the three initial 
conditions characterizing the solutions of the fixed function equation.
For this purpose we expand
$v(x)$ around $x = 0$ 
\be\label{xasmp}
v(x) = c \left( 1 + v_1 \, x + v_2 \, x^2 \right) + \cO(x^3) \, .
\ee
Substituting this ansatz into \eqref{vdgl} and expanding the resulting equation
at $x=0$ results in a Laurent series starting with $x^{-2}$. Requiring that the 
series coefficients of $x^{-2}$ and $x^{-1}$ vanish fixes the
unknown coefficients $v_1$, $v_2$ in terms of $c$
\be
v_1 = \frac{342}{67},\quad v_2 = \frac{90}{3551} \, \left( 376 + 19845 \, \pi^2 \, c \right) \, .
\ee
The parameter $c$ remains undetermined at this stage. In the next subsection,
this parameter will be constrained by demanding regularity of the solutions 
at $r_{\rm sing}$ and $r=0$.

\subsubsection*{Regularity at $r = r_{\rm sing}$}
At this stage, we have restricted the set of candidate fixed functions
to a one-parameter family of functions $v(x; c)$, depending on the 
initial condition $c$ implemented at $x=0$. In the next step, we apply a numerical shooting
method to extend the asymptotic solutions \eqref{xasmp} to the interval $x \in [0, x_{\rm sing}]$.

Owed to the first order pole in \eqref{vdgl}, we a priori expect that the solution for a generic value $c$ 
will become singular at $x = x_{\rm sing}$. Expanding the solution in a Taylor series at $x_{\rm sing}$
\be
v(x; c) = \sum_{n=0}^\infty \frac{1}{n!} \, a_n(c) \, (x-x_{\rm sing})^n \, , 
\ee
and substituting this ansatz into \eqref{vdgl}, the l.h.s.\ of the differential equation
remains regular while the r.h.s.\ generically has a first order pole. The cancellation 
of the residue then gives rise to an additional constraint on the admissible fixed functions.
The shooting method allows to phrase this constraint in terms of the free parameter $c$, fixed at $x=0$.
Based on these preliminary considerations, we expect that only a discrete set of the candidate solutions 
is able to pass $x_{\rm sing}$ without becoming singular, thus limiting the allowed values of $c$ to
a discrete set.

In practice, we construct $v(x; c)$ by numerically integrating \eqref{vdgl} up to  $x_{\rm term} = x_{\rm sing} - \epsilon$, applying a BDF-algorithm.
Varying $c$, this gives rise to a one-parameter family of solutions $v(x; c)$, defined on the interval $x \in [0, x_{\rm term}]$. Substituting
these solutions into \eqref{Ntilde}, we define
\be\label{deltafct}
\delta(c) \equiv \lim_{x \rightarrow x_{\rm sing}} \tilde{\cN}(v(x; c), v'(x; c), v''(x; c); x) \, . 
\ee
The vanishing of the residue at $x = x_{\rm sing}$ is implied by the zeros of $\delta(c)$. In other words,
solutions $v(x; c)$, which are regular at $x_{\rm sing}$ correspond to critical values $c_{\rm crit}$ where
$\delta(c_{\rm crit}) = 0$. 
\FIGURE[t]{
\qquad \qquad \qquad \qquad  \includegraphics[width=7cm]{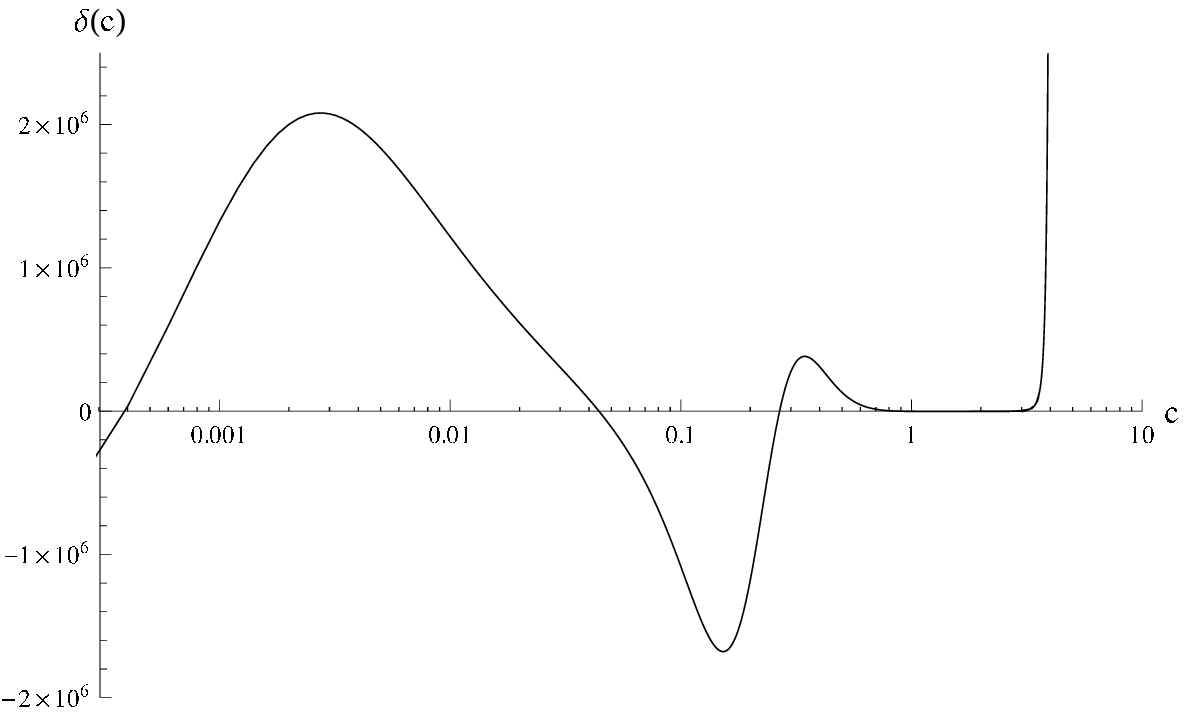} \qquad \qquad \qquad \qquad 
\caption{\label{fig.1} The function $\delta(c)$, eq.\ \eqref{deltafct}. Zeros of $\delta(c)$ correspond to solutions $v(x; c)$ which pass through the singularity at $r = r_{\rm sing}$ and can be continued to the patch $\cR_0$. For $c > 0$ there are three isolated zeros located at $c_{1,{\rm crit}} = 3.90 \times 10^{-4}$, $c_{2,{\rm crit}} = 4.41 \times 10^{-2}$, and $c_{3,{\rm crit}} = 0.255$. In addition there is one continuous window when $0.648 \le c \le 2.668$.}
}

The function $\delta(c)$ constructed via this algorithm is shown in Fig.\ \ref{fig.1}. Here the sampling of
$\delta(c)$ has been carried out in steps of $\Delta c = 10^{-4}$. The figure displays three
isolated zeros at
\be\label{ciso}
c_{1,{\rm crit}} = 3.90 \times 10^{-4} \, , \qquad c_{2,{\rm crit}} = 4.41 \times 10^{-2} \, , \qquad c_{3,{\rm crit}} = 0.255 \, ,
\ee
and a \emph{continuous window} of zeros located in the interval
\be\label{cwindow}
c_{-,{\rm crit}} \le c \le c_{+,{\rm crit}} \, , \qquad c_{-,{\rm crit}} = 0.648 \, , \quad c_{+,{\rm crit}} = 2.668 \, .
\ee
We also investigated the dependence of the critical values $c_{i,{\rm crit}}$ on the parameter $\epsilon$ and found that their position is stable, if $\epsilon$ is chosen sufficiently small.
When decreasing $\epsilon$, the function $\delta(c)$ is becoming steeper at the isolated critical points, while inside the window it is continuously approaching zero.

While the existence of isolated solutions has already been anticipated in the discussion above,
the appearance of the continuous window \eqref{cwindow} is somewhat surprising. In this case,
$\delta(c) = 0$ acts as an attractor for an entire interval of initial conditions. Since $\tilde{\cN}$ 
is build from $v(x)$ and its first and second derivatives, which balance at $x_{\rm sing}$
the solutions $v(x; c)$ in this interval are physically different also when continued to $\cR_0$, 
i.e., one does not ``lose memory'' by passing through this attractor. 

At this stage, the following remarks are in order. We have also investigated the behavior of $\delta(c)$ for negative $c$.
In this regime \eqref{deltafct} is positive definite, implying that these solutions cannot be extended into the patch $\cR_0$. Moreover,
by expanding \eqref{pdgl} for small values $c$, which essentially corresponds to linearizing \eqref{vdgl}, the solutions $v(x; c)$ can be found analytically. Substituting 
the analytic expression into $\delta(c)$ one can show that for $c < c_{1,{\rm crit}}$ there is only the trivial zero $c = 0$. In this regime the analytic and numerical studies match continuously, ensuring that there is no accumulation of zeros for $c \ll 1$.

\subsubsection*{Continuation to $r = 0$ and moving singularities}
In the final step, the candidate fixed functions encoded in \eqref{ciso} and \eqref{cwindow} are continued to the UV-patch $\cR_0$.
There are two potential obstacles, which may prevent the solutions to constitute valid fixed functions:
\begin{enumerate}
\item moving singularities at $r_{\rm term} > 0$.
\item the first order pole of \eqref{FFsolved} at $r=0$.
\end{enumerate}
Following the strategy of the previous subsection, we investigate these final conditions by numerically integrating \eqref{FFsolved} with initial conditions at $r_{\rm sing}$, employing an algorithm that takes the stiffness of the ODE at $r=0$ into account. Practically, we perform the continuation of $v(x; c_{\rm crit})$ to $\cR_0$ by first determining $v(c; c_{\rm crit})$ and its first and second derivative
at $x_{\rm sing} - \epsilon$. These values are then used as initial conditions for the profile function $y(r)$ at $r_{\rm sing} - \epsilon$, taking 
the proper Jacobians into account. Typically, we work with $\epsilon = 10^{-7}$, and we have checked that our results are insensitive to $\epsilon$ as long as it is sufficiently (but not too) small.

The only case in which we encounter a moving singularity is the solution emanating from $c_{3, {\rm crit}}$. The corresponding numerical integration is shown in Fig.\ \ref{fig.5}. 
This solution develops a logarithmic singularity of the form \eqref{logsing} at $r_{\rm term} = 1.26$ and can therefore not be completed in a regular solution extending down to $r=0$.
Notably, the other candidate fixed functions \eqref{ciso} and in particular \emph{all} solutions from the continuous window \eqref{cwindow} are free from this type of singularities.
\FIGURE[t]{
\qquad \qquad \includegraphics[width=7cm]{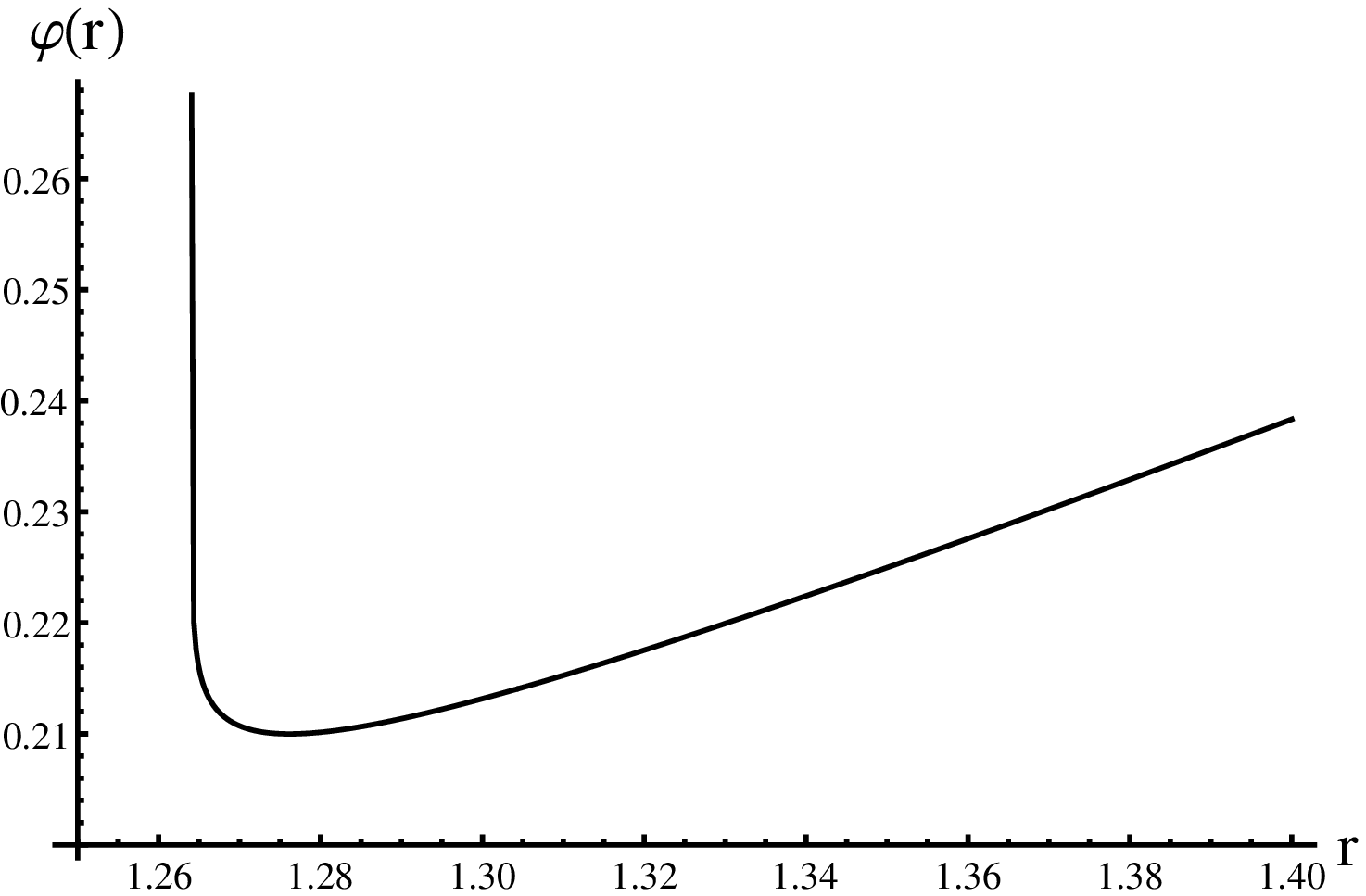} \qquad \qquad 
\caption{\label{fig.5} Numerical continuation of the candidate fixed function $c_{3,{\rm crit}} = 0.255$ to the IR-patch $\cR_0$. The solution terminates
in a moving singularity at $r_{\rm term} = 1.26$ and cannot be continued to $r=0$.}
}

Finally, we encounter the fixed singularity at $r=0$. In analogy to the singularity at $r_{\rm sing}$, we require that this pole in \eqref{fpeq} is compensated by a zero 
in the numerator \eqref{Nfct}. We then substitute our numerical solutions into $\cN(\va, \va', \va'', r)$. For the isolated solutions $c_{1, {\rm crit}}$ and $c_{2, {\rm crit}}$ and the two boundaries
of the continuous window, the result is shown in the left and the right panel of Fig.\ \ref{fig.2}, respectively. 
At $r=0$, $\cN(\va, \va', \va'', r; c_{i, {\rm crit}})$ vanishes, establishing that the last condition is satisfied automatically for all remaining fixed functions.
\FIGURE[t]{
\includegraphics[width=6cm]{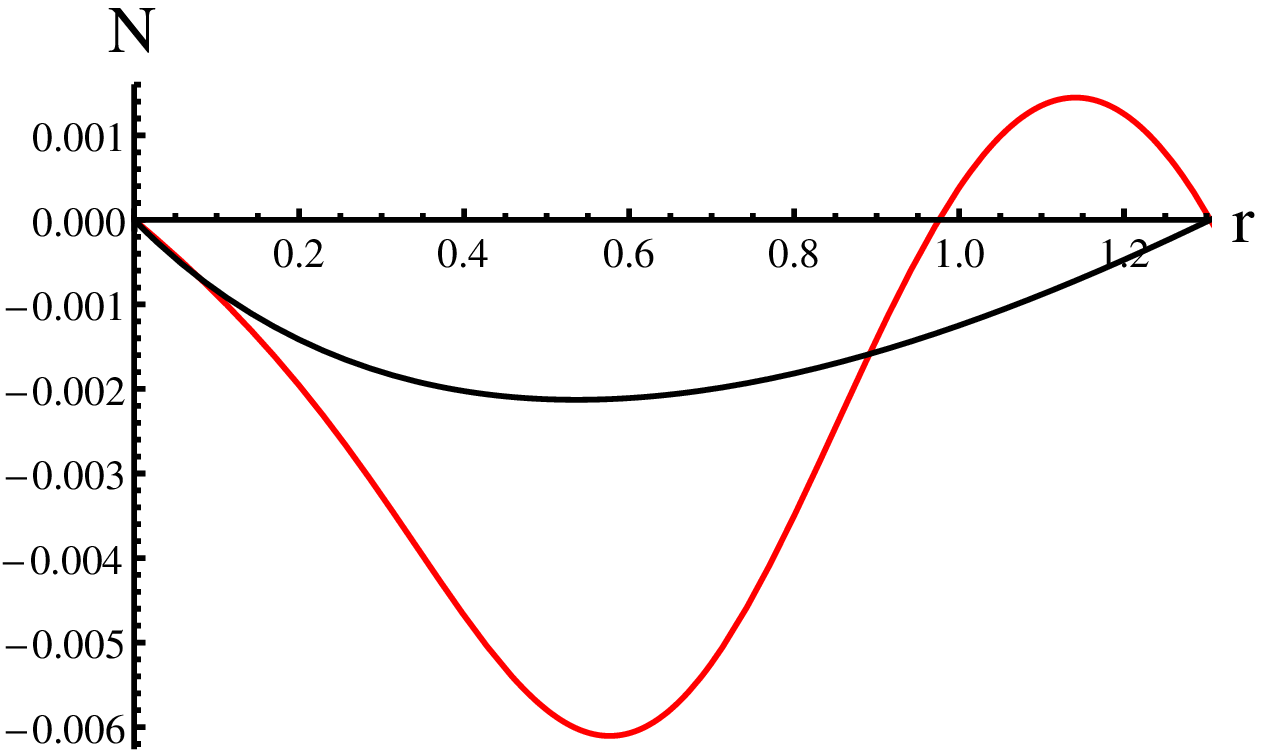} \; \; \; 
\includegraphics[width=6cm]{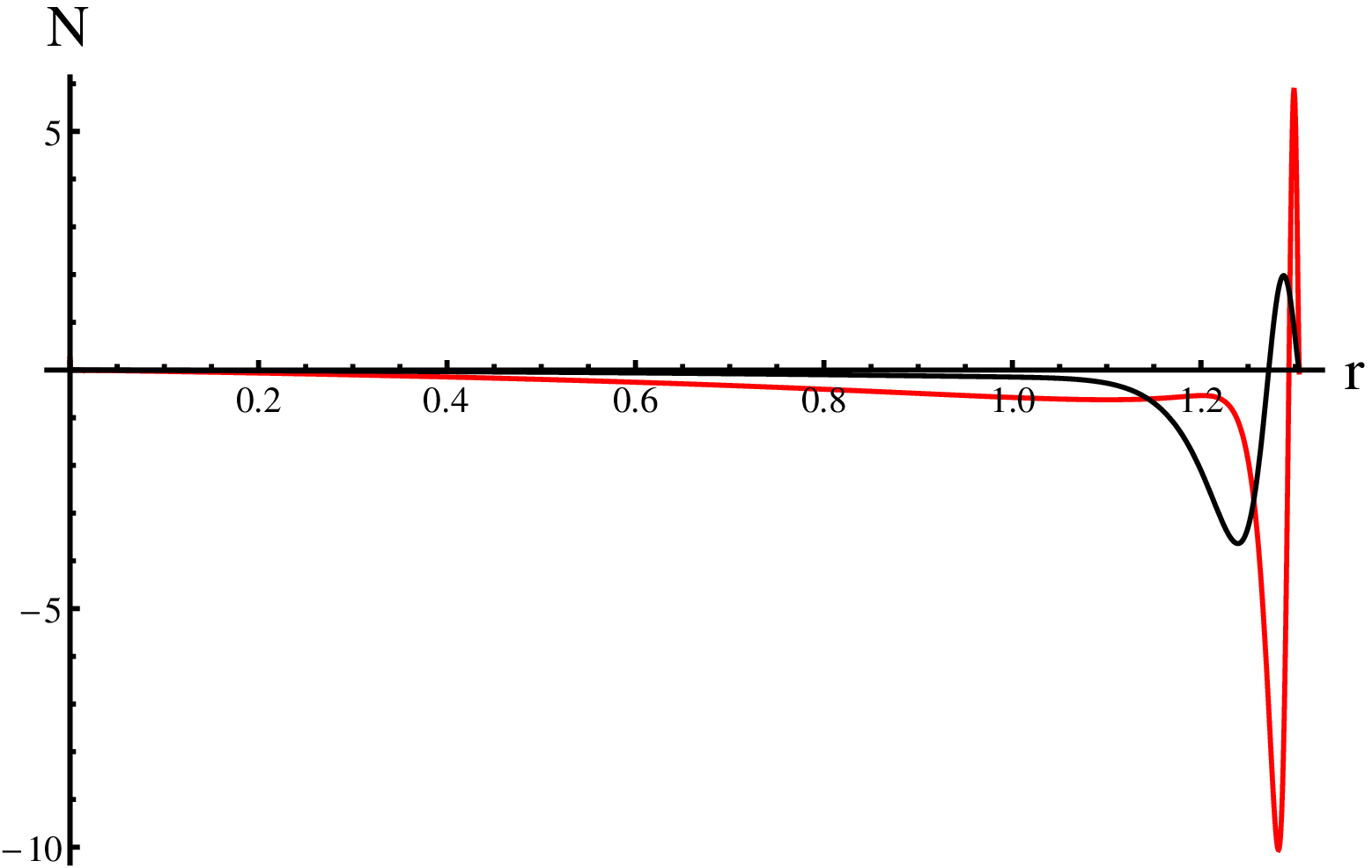} 
\caption{\label{fig.2} The function $\cN$ evaluated along the critical solutions $c_{1, {\rm crit}}$ (black curve) and $c_{2, {\rm crit}}$ (red curve)
and the boundaries of the continuous window $c_{-, {\rm crit}}$ (black curve) and $c_{+, {\rm crit}}$ (red curve) are shown in the left and right panel, respectively. For presentational purposes,
$\cN(r; c_{1,{\rm crit}})$ has been magnified by a factor $10^3$. The vanishing of the numerator at $r=0$ ensures that the solutions are regular on the entire interval $0 \le r < r_{\rm sing}$.}}

In summary, we have established that besides the Gaussian fixed function \eqref{GFP},
the non-linear fixed function equation \eqref{fpeq} gives rise to two isolated
and one continuous family of fixed functions which are regular on the entire positive real axis $r \ge 0$.
They can conveniently be characterized by their asymptotics in the IR ($r \gg 1$):
\be\label{ciso2}
c_{1,{\rm crit}} = 3.90 \times 10^{-4} \, , \qquad c_{2,{\rm crit}} = 4.41 \times 10^{-2} \, ,
\ee
and
\be\label{cwindow2}
c_{-,{\rm crit}} \le c \le c_{+,{\rm crit}} \, , \qquad c_{-,{\rm crit}} = 0.648 \, , \quad c_{+,{\rm crit}} = 2.668 \, .
\ee
The existence of these fixed functions is the central result of this section and we will continue by studying their properties in the remainder of this work.

\section{Properties of the fixed functions}
\label{sect.4}
As the central result obtained so far, we established that
the flow equation \eqref{pdgl} of conformally reduced QEG
in $d=3$ possesses two isolated and one continuum
family of regular fixed functions. The characteristic
properties of these solutions will be discussed in this section.
\subsection{Characteristics of the numerical solutions}
We start by displaying the regular fixed functions $\va_{i,*}(r)$ arising
from the critical values $c_{i, {\rm crit}}$, eqs.\ \eqref{ciso2} and \eqref{cwindow2},
in Fig.\ \ref{fig.3}. The corresponding profile functions $y(r)$
introduced in \eqref{varedef} are shown in the right diagrams.
They interpolate continuously between the UV ($r \ll 1$) and the IR ($r \gg 1$)-regime\footnote{In a slight abuse of language we will refer to the expansions of our fixed functions for large and small values $k$ as UV and IR. The corresponding fixed functional $\Gamma_\ast$ is of course independent of $k$.},
where they become constant.
The crossover occurs when the RG-scale is of the same order of magnitude as the background curvature, i.e., $r = R/k^2 \approx 1$.
\FIGURE[t]{
\includegraphics[width=6.5cm]{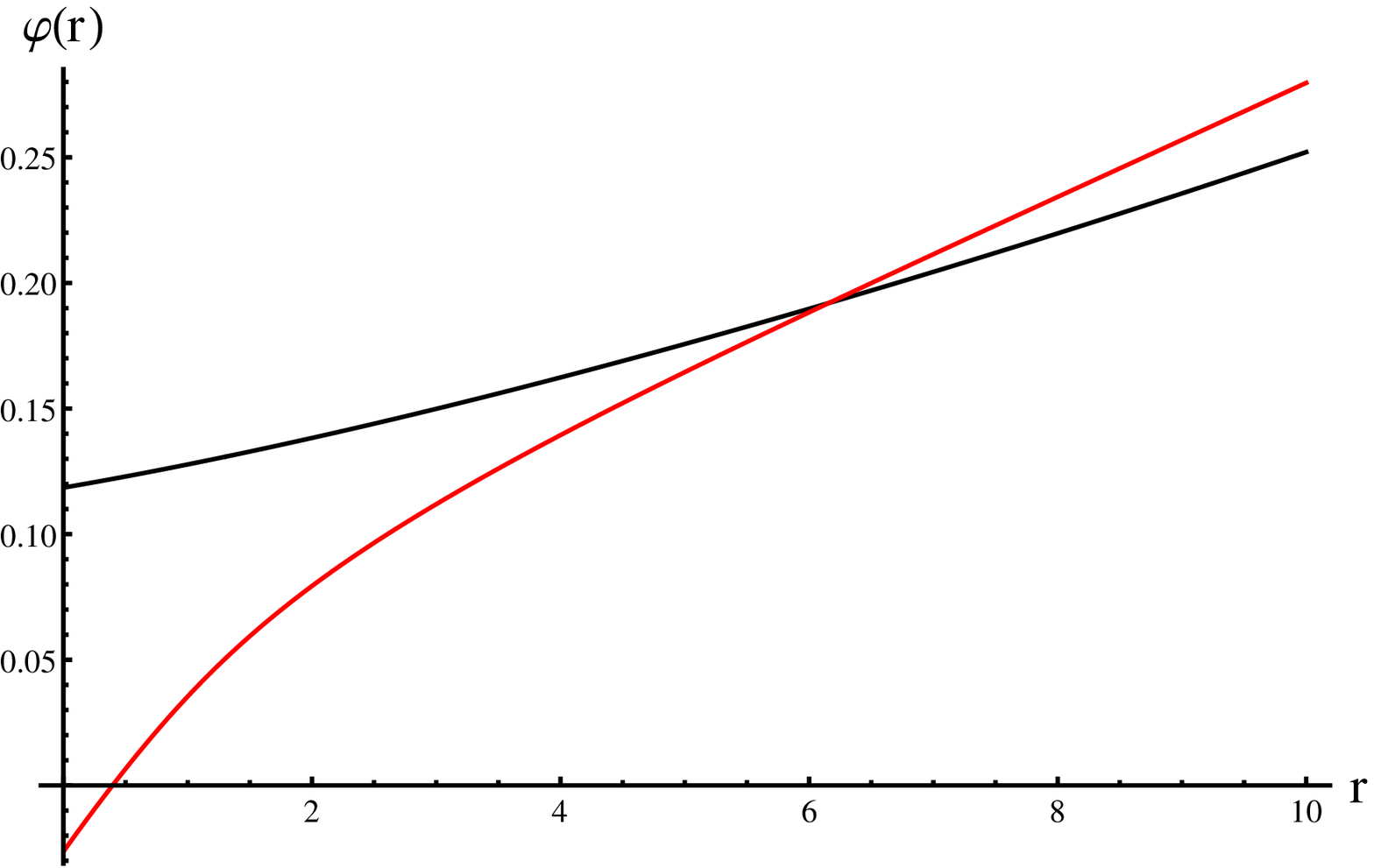} \qquad
\includegraphics[width=6.5cm]{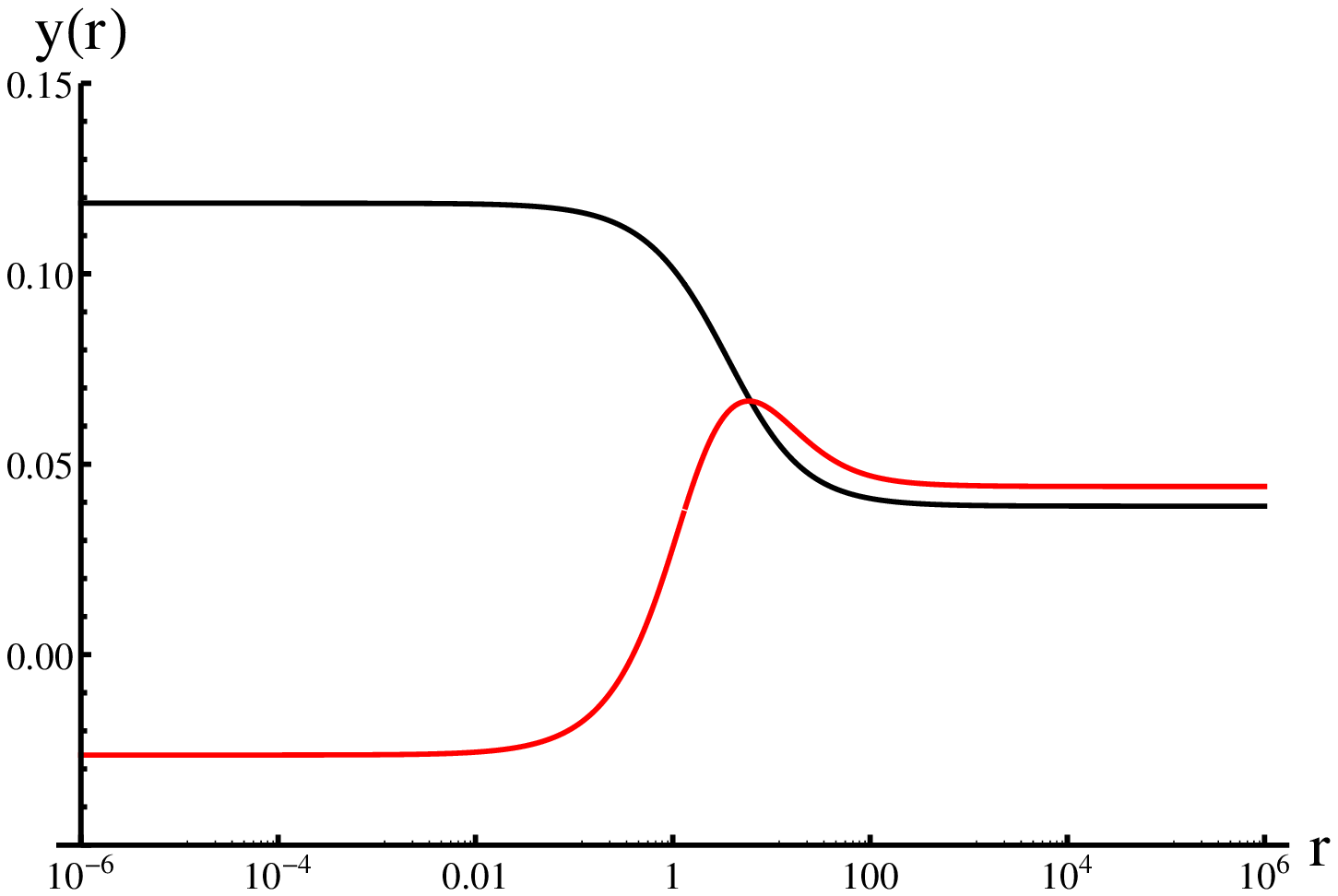} \\ 
\includegraphics[width=6.3cm]{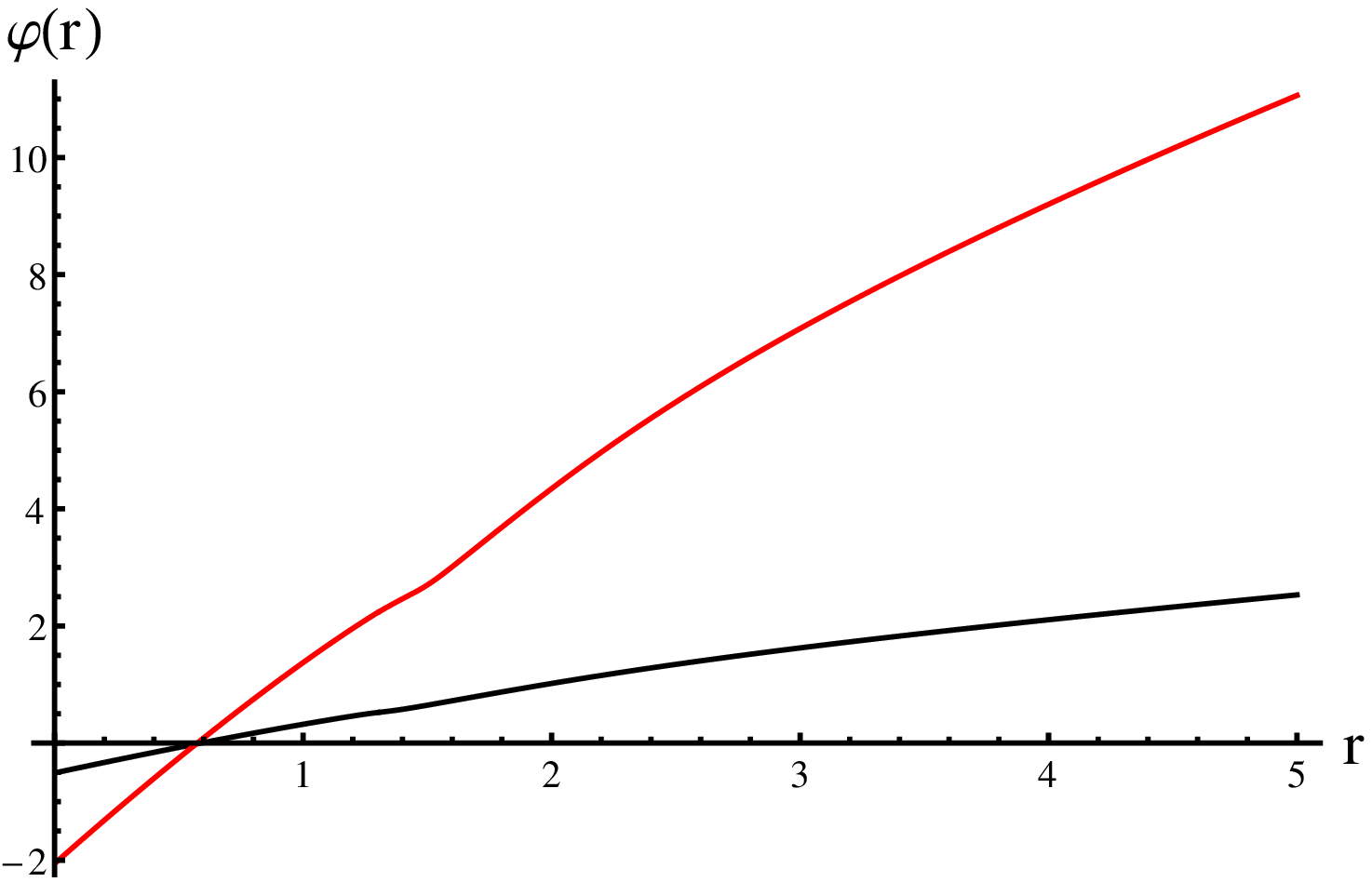} \qquad
\includegraphics[width=6.3cm]{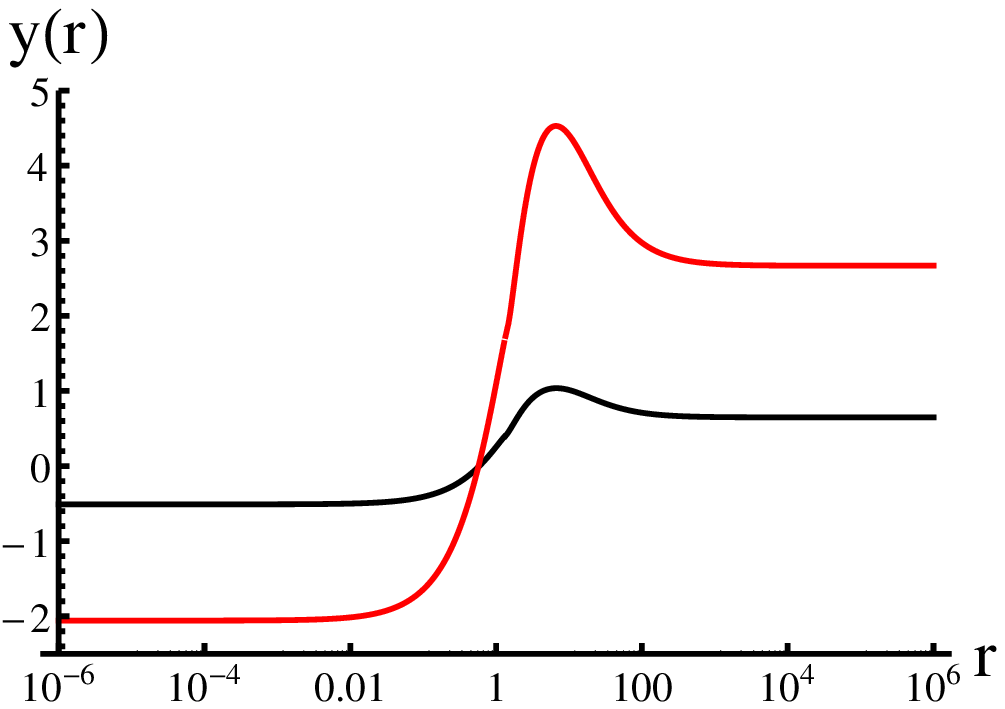}
\caption{\label{fig.3} All complete, regular fixed functions \eqref{ciso2} and \eqref{cwindow2}. The upper panels show
the isolated fixed functions characterized by $c_{1, {\rm crit}}$ (black) and $c_{2, {\rm crit}}$ (red).
The lower panels display the boundary solutions of the continuous window $c_{-, {\rm crit}}$ (black) and $c_{+, {\rm crit}}$ (red). For presentational purposes the $c_{1, {\rm crit}}$-solution has been multiplied by a factor $100$.}
}

The fixed functions $\va_{i,*}(r)$ resulting from these profile functions are shown in the left diagrams of Fig.\ \ref{fig.3}.
They are all bounded from below and increase monotonically with increasing $r$. The asymptotic behavior of the profile functions
furthermore implies that the $\va_{i,*}(r)$ interpolate continuously between
\be\label{vainter}
\begin{array}{ll}
\va_*(r) \sim u_{i,0}^* \, , \qquad & r \ll 1 \; \; {\rm (UV)} \\[1.2ex]
\va_*(r) \sim 6^{-3/2} c_{i, {\rm crit}} r^{3/2} \, , \qquad \qquad & r \gg 1 \; \; {\rm (IR)} \, .
\end{array}
\ee 
Here the IR-coefficients $c_{i, {\rm crit}}$ are given in eqs.\ \eqref{ciso2} and \eqref{cwindow2},
while the constants $u_{i,0}^*$ determining the UV-asymptotics are listed in the first line
of Table \ref{t.1}. When comparing the four solutions, there is an obvious qualitative
difference between $\va_{1,*}(r)$ and the other fixed functions:
$\va_{1,*}(r)$ is singled out
by the fact that it is the only solution that remains positive in the UV. Thus it is
the only fixed function that remains positive definite throughout.

At this stage it is illustrative to restore the dimensionful quantities and investigate the asymptotic behavior of the conformal field theories associated with the fixed functions in the IR. 
Substituting the IR-asymptotics \eqref{vainter} into \eqref{dimless1} and restoring the spacetime integral
via \eqref{eq:ansatz}, the effective action $\Gamma$ becomes
\be\label{IRasymp}
\Gamma_* = \frac{1}{6^{3/2}} \int d^3x \sqrt{g} \, c_{i,{\rm crit}} \, R^{3/2} \, . 
\ee
The effective action resulting from our fixed functions is precisely given by
the power-counting marginal operator of the theory with the coupling
constant determined by $c_{i,{\rm crit}}$. Notably, it has the same form as
the one-loop effective action \cite{Avramidi:2000bm}. A similar structure for
the IR-asymptotics of $\Gamma_*$ has also been observed
in four dimensions \cite{Benedetti:2012dx}. In this case it was, however, not possible to obtain the 
allowed values $c_{i,{\rm crit}}$ since the complete regular solutions of the differential equation describing the fixed points has not been constructed.

We stress that none of the properties discussed here are put in ``by hand''. They all arise
from solving the fixed function equation \eqref{fpeq} and this have the status of 
predictions from the quantum theory.

\subsection{Asymptotic expansions}
\label{sect4.2}
Refs.\ \cite{Codello:2007bd,Machado:2007ea,Codello:2008vh,Rahmede:2011zz} studied the RG-flow of $f(R)$-gravity by implementing 
a polynomial ansatz for the function $f(R)$ and subsequently analyzing the properties of the resulting NGFP.\footnote{For a similar study in scalar-tensor theories see \cite{Narain:2009gb}.}
It is thus instructive to compare the full numerical solutions shown in Fig.\ \ref{fig.3} 
and their polynomial approximation found by solving the fixed point equation recursively.
\subsubsection*{Polynomial expansion for small $r$}
In order to capture the asymptotic behavior of the fixed functions in the UV, we expand the fixed function
in a Taylor series at $r=0$ up to a fixed order $N$
\be\label{expr0}
\va_*(r) = \sum_{n=0}^N \, u_n^* \, r^n \, . 
\ee
\TABLE[t]{
\begin{tabular}{|c||c|c|c|c|}
\hline
 &  $\qquad c_{1,{\rm crit}} \qquad$  & $\qquad c_{2,{\rm crit}} \qquad$ &  $\qquad c_{-,{\rm crit}} \qquad$  &  $\qquad c_{+,{\rm crit}} \qquad$  \\ \hline \hline
 $\qquad u_0^* \qquad$  & $1.185\times 10^{-3}$  & $-2.635\times 10^{-2}$ & $-5.099\times 10^{-1}$ & $-2.056$\\\hline
 $u_1^*$                & $8.466 \times 10^{-5}$ & $6.917\times10^{-2}$   & $8.211\times 10^{-1}$  & $2.996$\\\hline\hline
 $u_{2, {\rm fit}}^*$   & $8.593 \times 10^{-6}$ & $-5.697\times 10^{-3}$ & $1.349$                & $10.554$\\
 $u_{2, {\rm rec}}^*$   & $8.595\times 10^{-6}$  & $-6.143\times 10^{-3}$ & $-5.515\times 10^{-2}$ & $-0.182$\\\hline
 $u_{3,{\rm fit}}^*$    & $-1.080\times 10^{-6}$ & $-3.922\times 10^{-3}$ & $-8.109$               & $-56.701$\\
 $u_{3,{\rm rec}}^*$    & $-1.098\times 10^{-6}$ & $-8.828\times 10^{-4}$ & $-5.389\times 10^{-3}$ & $-1.450\times 10^{-2}$\\\hline
 $u_{4,{\rm fit}}^*$    & $2.944\times 10^{-7}$  & $9.770\times 10^{-3}$  & $22.874$               & $149.672$\\
 $u_{4,{\rm rec}}^*$    & $4.104\times 10^{-7}$  & $3.246\times 10^{-4}$  & $-1.549\times 10^{-3}$ & $-3.549\times 10^{-3}$\\\hline
 $u_{5,{\rm fit}}^*$    & $-1.222\times 10^{-8}$ & $-1.874\times 10^{-2}$ & $-37.364$              & $-230.235$\\
 $u_{5,{\rm rec}}^*$    & $2.474\times 10^{-8}$  & $-1.566\times10^{-4}$ & $-6.187\times 10 ^{-4}$ & $-1.2289\times 10^{-3}$\\\hline
\end{tabular}
\caption{\label{t.1} Series coefficients appearing in the expansion \eqref{expr0} of the fixed functions $\va_{i,*}(r)$.
The two free parameters $u^*_{i,0}$ and $u^*_{i,1}$ characterize the fixed functions in the UV-regime for $r \ll 1$. Together with the higher order
momenta $u_{n, {\rm fit}}^*$ they are obtained by fitting the polynomial \eqref{expr0} to the numerical solutions $\va_{i,*}(r)$,
while the values $u_{n, {\rm rec}}^*$ are determined recursively by solving the system of linear equations arising from expanding \eqref{fpeq} up to order $r^{n-2}$.
}}
Substituting this ansatz, the expansion of  \eqref{fpeq} at $r=0$
gives rise to a first order pole. Following the discussion of Section \ref{sect:3.3}, this pole fixes one of the three free parameters
characterizing the fixed functions. The remaining two constants of integration
can then be taken as $u_0^*$ and $u_1^*$. By solving the expanded fixed function equation order by order in $r$, 
the coefficients $u_n$ for $n \ge 2$ can 
then be found recursively as functions of these free parameters, i.e., the coefficient $u_2^*$ is fixed by a condition at order $r^0$, etc.
It thereby turns out that the resulting system of equations has a triangular structure: the first 
equation is linear in $u_2^*$ and thus gives a unique value for $u_2^*$ in terms of $u_0^*$ and $u_1^*$.
The next order equation is again linear in $u_3^*$, and uniquely determines $u_3^*$ in 
terms of the lower order couplings, and so on. This structure guarantees that the solutions are characterized uniquely by $u_0^*$ and $u_1^*$.
 
With the complete numerical solutions at our disposal, we can fit the polynomials \eqref{expr0} (with $N=50$) to the fixed functions $\va_{i,*}(r)$ restricted to the
UV-patch $\cR_0$. The parameters $u_0^*$ and $u_1^*$, which characterize the solutions, and the first four subsequent coefficients $u_{i,{\rm fit}}^*$ 
are shown in Table \ref{t.1}. 
Taking the parameters $u_{i,0}^*$ and $u_{i,1}^*$ (the first index labels the various solutions $c_{i,{\rm crit}}$) as initial conditions for the polynomially expanded fixed function equation,
we also determine the higher order expansion coefficients recursively. These are summerized by the entries
$u_{i,{\rm rec}}^*$ of Table \ref{t.1}. Comparing the expansions obtained in these two ways, we 
observe that the expansion coefficients of $\va_{1,*}(r)$ show a reasonable agreement, while
the expansions of the other solutions already show some level of disagreement at order $u_2^*$.
 
These observations can be formalized by studying the convergence radius of the recursive solutions $r_{\rm con}$.
We determine $r_{\rm con}$ via the quotient criterion
\be\label{qcrit}
r_{\rm con} = \lim_{n \rightarrow \infty} \frac{|u_n|}{|u_{n+1}|} \, .
\ee
For the fixed function $\va_{1,*}(r)$ the expansion coefficients $u_n$
are easily obtained up to order 200. The resulting first 200
terms in the series \eqref{qcrit} are then shown in the left diagram
of Fig.\ \ref{fig.6}. For $n>50$ the series converges rapidly.
Fitting a Laurent series including a constant, $r^{-1}$ and $r^{-2}$ term, we obtain
\be
r_{\rm con} = 1.30707 \approx r_{\rm sing} \, .
\ee
Thus the polynomial expansion for $\va_{1,*}(r)$ converges up to the singularity $r_{\rm sing}$.
This result is also demonstrated in the right panel of Fig.\ \ref{fig.6}, which displays 
the direct comparison between the fixed function obtained numerically (black) and the polynomial approximation (red). 
\FIGURE[t]{
\includegraphics[width=6.5cm]{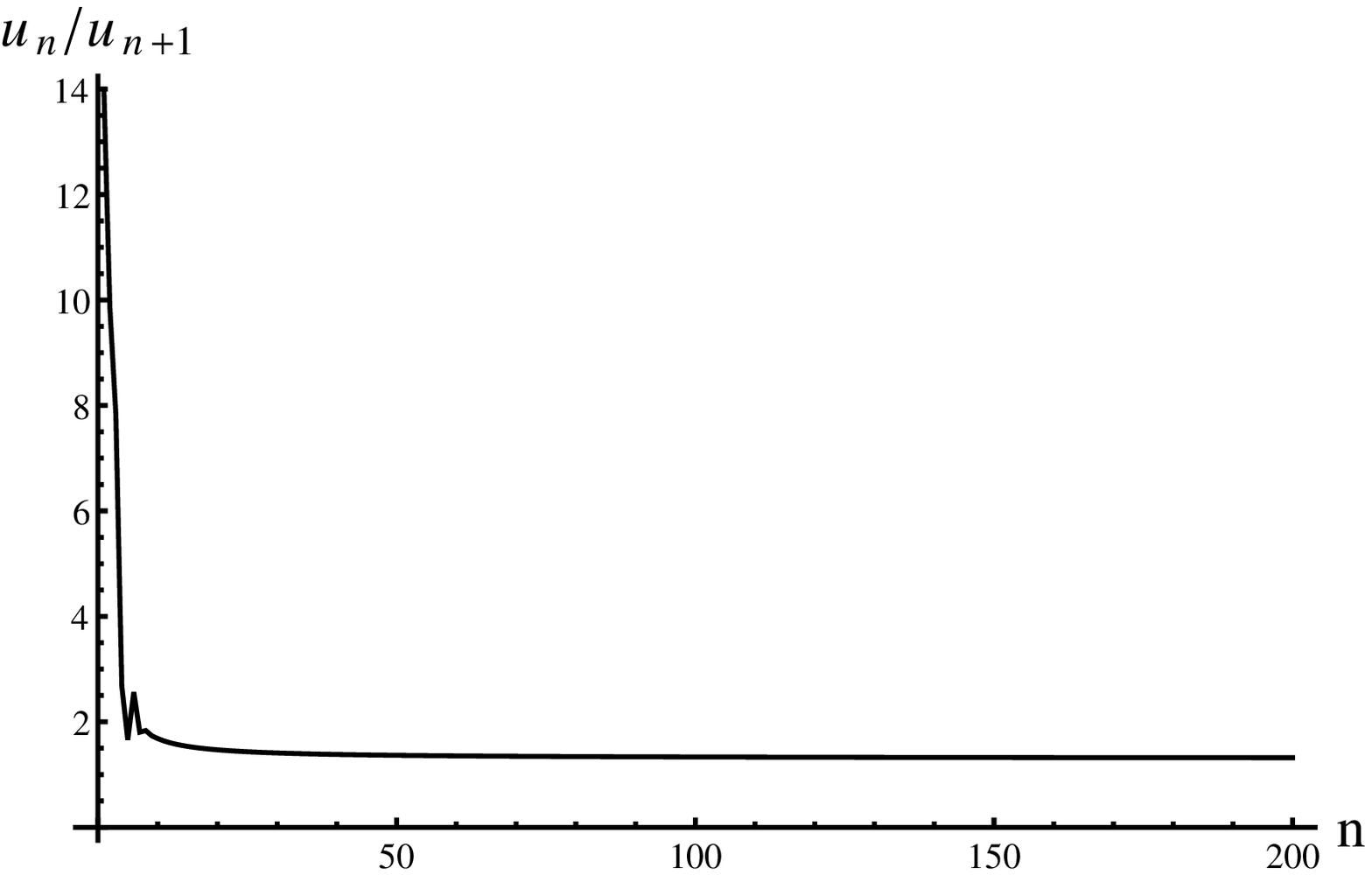} \, 
\includegraphics[width=6.5cm]{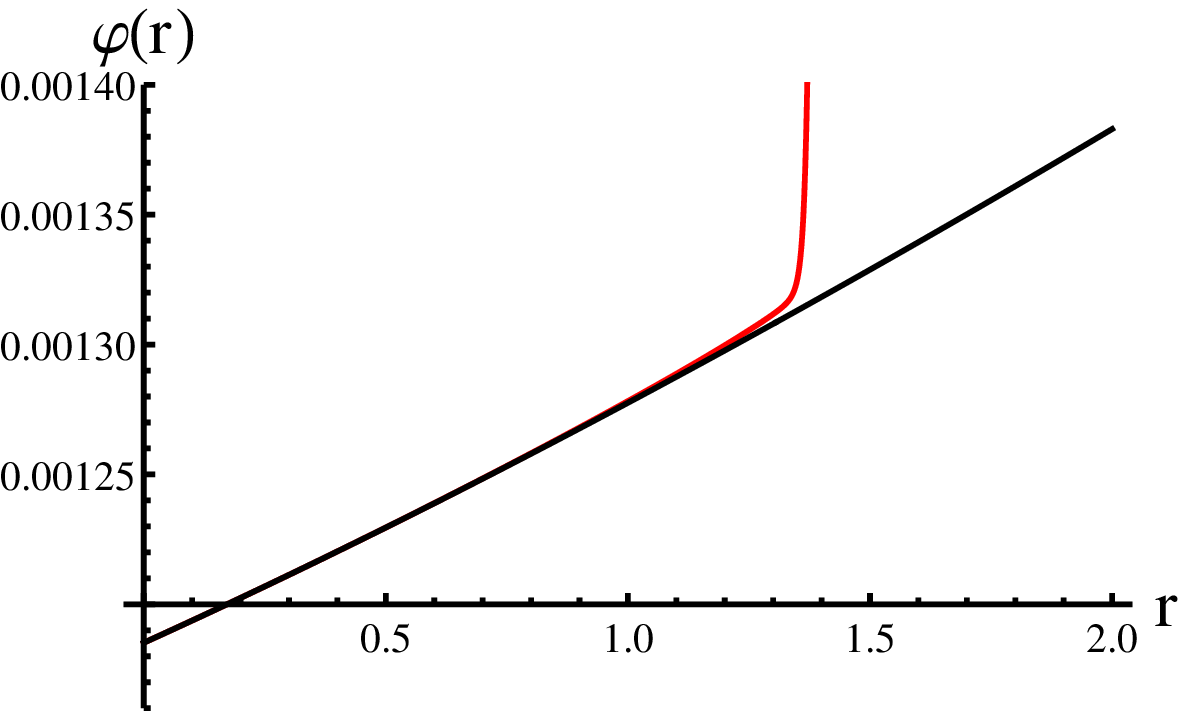} \, 
\caption{\label{fig.6} Series coefficients \eqref{qcrit}, arising from the recursive solution of the fixed function equation
with initial conditions $u_{1,0}^*$ and $u_{1,1}^*$. The series converges up to $r_{\rm con} = r_{\rm sing}$.
This is confirmed in the right panel which compares $\va_{1,*}(r)$ obtained numerically (black) and recursively (red).}}

Applying the quotient criterion to the other fixed functions quickly reveals that their radius of convergence is actually zero.
We exemplify this result in Fig.\ \ref{fig.7}, which compares the numerical solution $\va_{2,*}(r)$
and the corresponding polynomial approximation with initial values $u_{2,0}^*$ and $u_{2,1}^*$.
\emph{Increasing} the order of the Taylor-expansion systematically actually \emph{decreases} the quality of the approximation.
This behavior is prototypical for the non-convergent nature of the series. 
\FIGURE[t]{
\qquad \qquad \includegraphics[width=7cm]{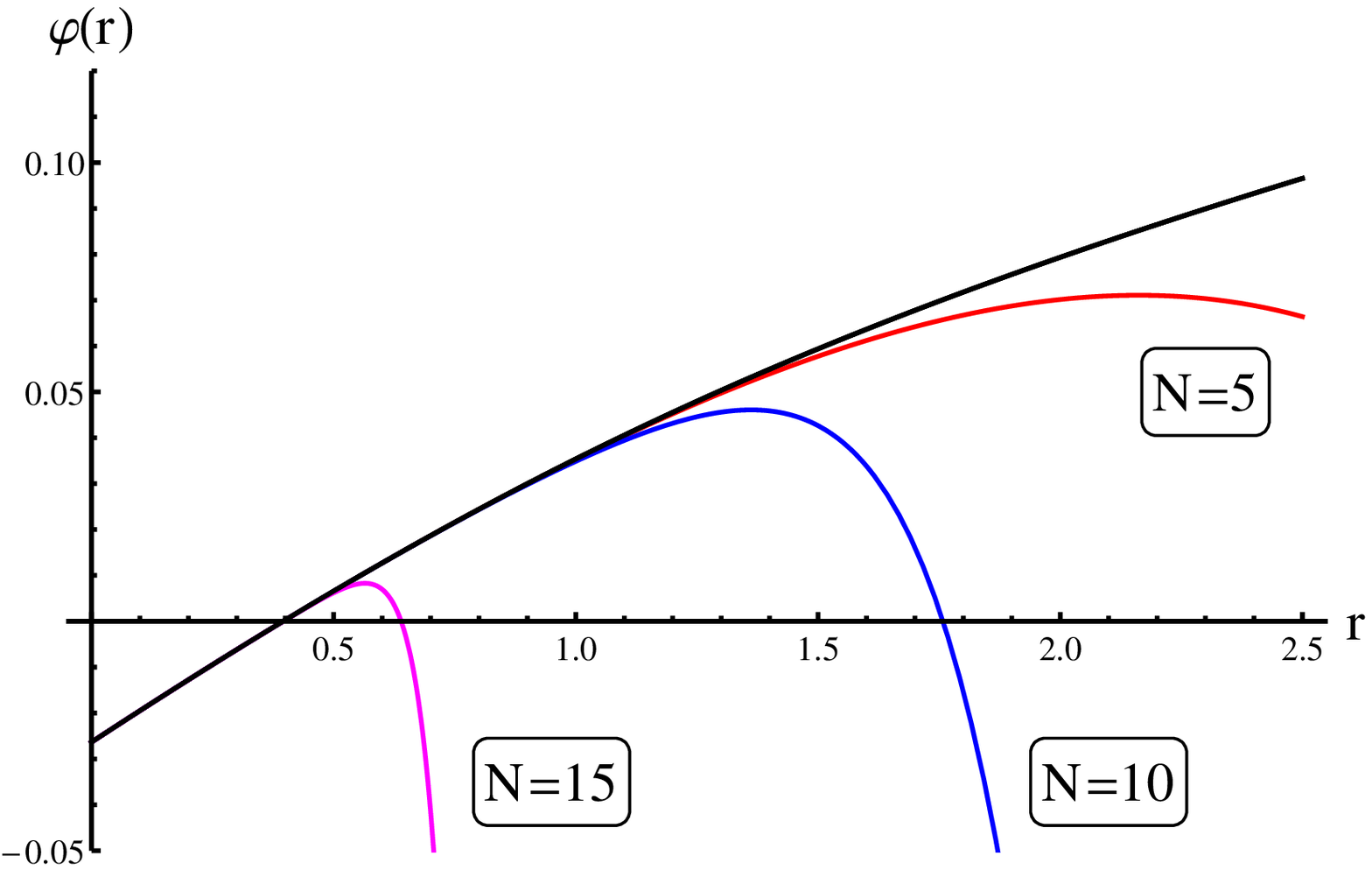} \qquad \qquad 
\caption{\label{fig.7} Comparison between the fixed function $\va_{2,*}(r)$ obtained numerically (black)
and its polynomial approximation \eqref{expr0} truncated at order $N=5$ (red), $N=10$ (blue), and $N=15$ (magenta).
Notably the quality of the polynomial approximation \emph{decreases} with increasing order of the approximation,
indicating that its radius of convergence \eqref{qcrit} is actually zero.}
}

\subsubsection*{Physical coupling constants in the UV}
The UV-behavior of the fixed functions is governed by the expansion \eqref{expr0} around $r=0$. 
Restoring dimensionful quantities, the UV-asymptotics of the average action $\Gamma_*[g]$ resulting
from the fixed functions $\va_{i,*}(r)$ has the curvature expansion
\be\label{UVexp}
\lim_{k \rightarrow \infty} \, \Gamma_*[g] = \int d^3x \sqrt{g} \left[ \frac{1}{16 \pi G_N} \left( -R + 2 \Lambda_k \right) + \bar{u}_2 R^2 + \ldots \right] \, , 
\ee
where $\Lambda$, $G_N$ and $\bar{u}_2$ are the \emph{dimensionful} cosmological constant, Newtons constant and $R^2$-coupling, respectively.
They are related to their dimensionless counterparts by
\be
\Lambda_k = \lambda_k k^2 \, , \qquad G_k = g_k k^{-1} \, , \qquad \bar{u}_2 = u_2 k \, .
\ee
A comparison between the polynomial expansion \eqref{expr0} and the Einstein-Hilbert action allows to express the dimensionless physical couplings
to the expansion coefficients
\be\label{dimlessgl}
\lambda_* = - \frac{u_0^*}{2 u_1^*} \, , \qquad g_* = - \frac{1}{16 \pi u_1^*} \, . 
\ee

Notably, the structure of \eqref{UVexp} closely
resembles the new massive gravity \cite{Bergshoeff:2009hq,Bergshoeff:2009aq}.\footnote{For a treatment in the framework of the FRGE see \cite{Ohta:2012vb}.}
In order to fully exploit this connection, it is useful to trade the coupling $u_2$ for the mass of the massive gravity mode. This relation is found as follows.
Substituting the $f_k(R)$ resulting from the expansion \eqref{UVexp} into \eqref{G2var} and afterwards taking the flat-space limit by setting $R = 0, \Lambda = 0$, we obtain
\be\label{mass}
\cK|_{\rm flat \, space} = \frac{2(d-1)^2}{d^2} \bar{u}_2 \, p^2 \, \left( p^2 + \frac{(d-2)}{4(d-1)} \, \frac{\bar{u}_1}{\bar{u}_2} \right) \, . 
\ee
Performing a Wick-rotation the zero in \eqref{mass} yields the mass of the ``massive graviton'' in $d=3$
\be\label{mgmass}
\bar{m}_0^2 = \frac{\bar{u}_1}{8 \bar{u}_2} =   \frac{u_1^*}{8 u_2^*} k^2 \equiv m^2_* \, k^2 \, .
\ee

\TABLE[t]{
\begin{tabular}{||c||c||c|c|c||}
                 &  $c_{i,\rm crit}$      & $\lambda_*$ & $g_*$ & $m^2_*$ \\ \hline \hline
$c_{1,\rm crit}$ &  $3.90 \times 10^{-4}$ & $-7.00$ & $-235$ & $1.23$ \\ 
$c_{2,\rm crit}$ &  $4.41 \times 10^{-2}$ & $0.19$ & $- 0.29$ & $-8.6 \times 10^{-3}$ \\ \hline
$c_{-,\rm crit}$ &  $0.648$ &  $0.31$ & $- 2.4 \times 10^{-2}$ & $7.6 \times 10^{-2}$ \\
$c_{+,\rm crit}$ &  $2.668$ & $0.34$ & $- 6.6 \times 10^{-3}$ & $3.5 \times 10^{-2}$  \\  \hline
\end{tabular}
\caption{\label{tab.4} Physical couplings arising from the polynomial expansion of the fixed functions $\va_{i,*}(r)$ in the UV. The values are obtained by substituting
the coefficients $u_{n, {\rm fit}}^*$ into the relations \eqref{dimlessgl} and \eqref{mgmass}.
}}
The values of the physical dimensionless couplings \eqref{dimlessgl} and \eqref{mgmass} arising from the polynomial expansion of our fixed functions in the UV
are collected in the third, fourth, and fifth column of Table \ref{tab.4}. In addition their values across the continuous window of fixed functions \eqref{cwindow2} 
are displayed in Fig.\ \ref{fig.4}.

We find that the the mass of the massive graviton is always positive
while Newtons constant remains negative throughout. While the later result looks unphysical at first sight, one should keep in mind though that the kinetic term of the conformal factor (the latter driving all the quantum effects considered in this paper) actually comes with the ``wrong sign'' for a kinetic term. This is precisely compensated by the negative value of Newtons constant, so that the fixed function produces a kinetic term that is positive definite.

\subsection{Linear perturbations and relevant operators}
\label{sect:4.3}
An additional characteristic of the fixed functions $\va_*(r)$ are their
UV-relevant deformations, which will be studied in this subsection.
For this purpose, we return to the full, $t$-dependent partial differential equation \eqref{pdgl}
and substitute the ansatz
\be\label{defans}
\va(t, r) = \va_*(r) + \epsilon \, e^{-\theta t} \, v(r) \, . 
\ee
Here $\theta$ denotes the stability coefficient associated with the perturbation $v(r)$,
$\va_*(r)$ specifies the fixed function for which we study the deformations,
and $\epsilon$ is a bookkeeping parameter, indicating that the deformations
under consideration are actual infinitesimal. By definition, UV-relevant deformations approach $\va_*(r)$ as $t \rightarrow \infty$
and are thus characterized by $\theta > 0$.
\FIGURE[t!]{
\includegraphics[width=4.5cm]{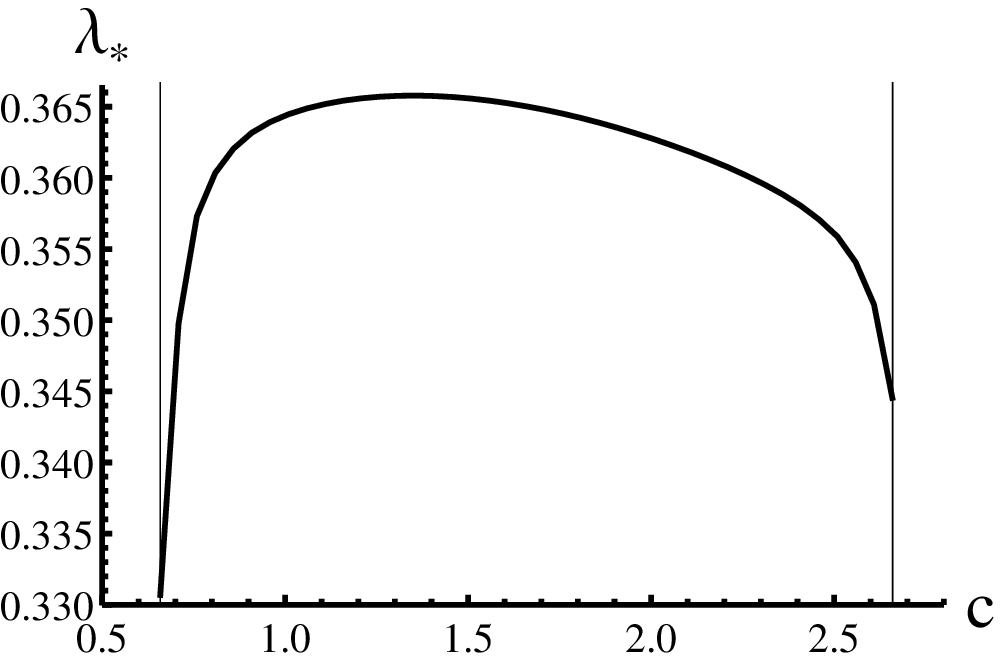} \, 
\includegraphics[width=4.5cm]{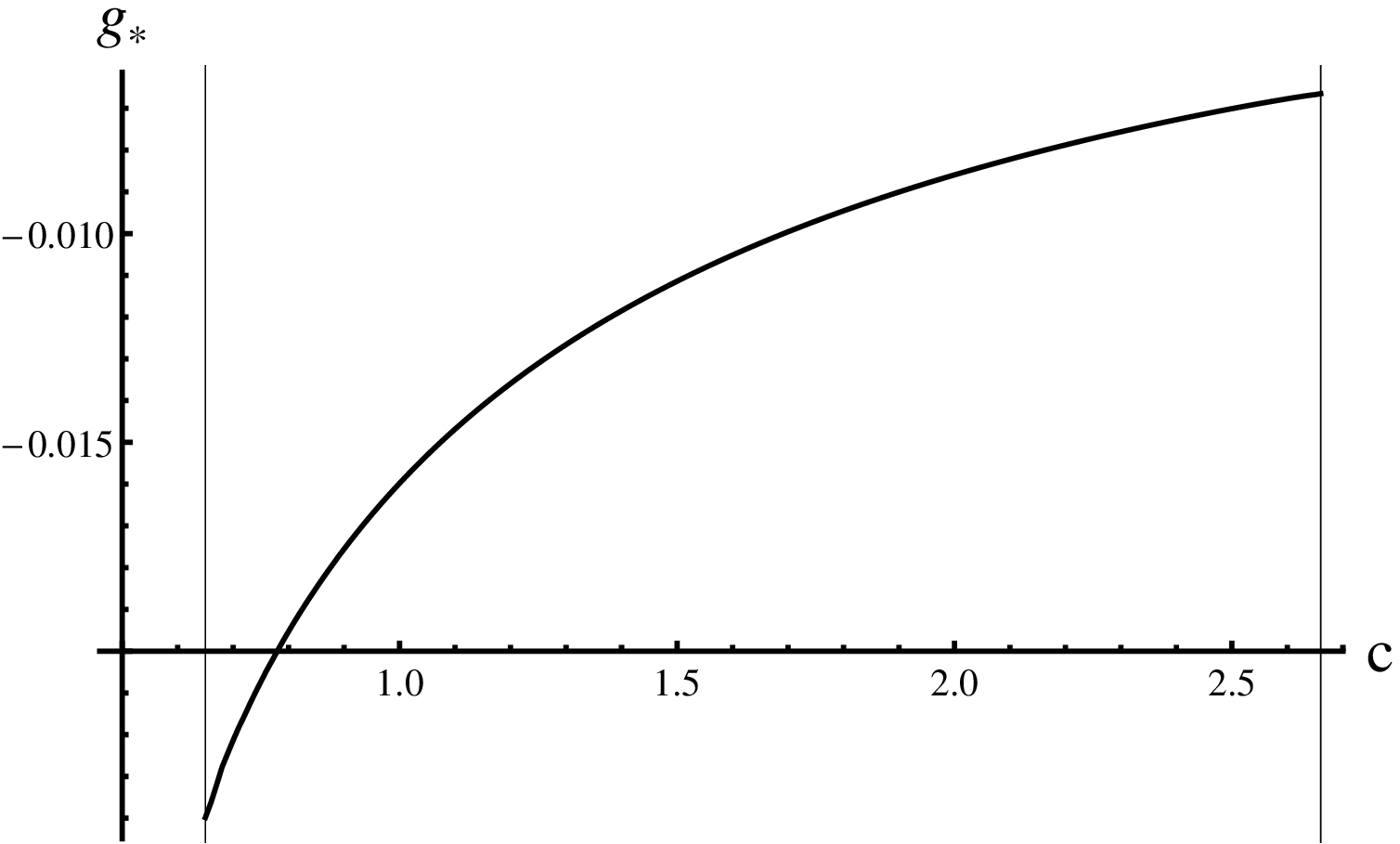} \, 
\includegraphics[width=4.5cm]{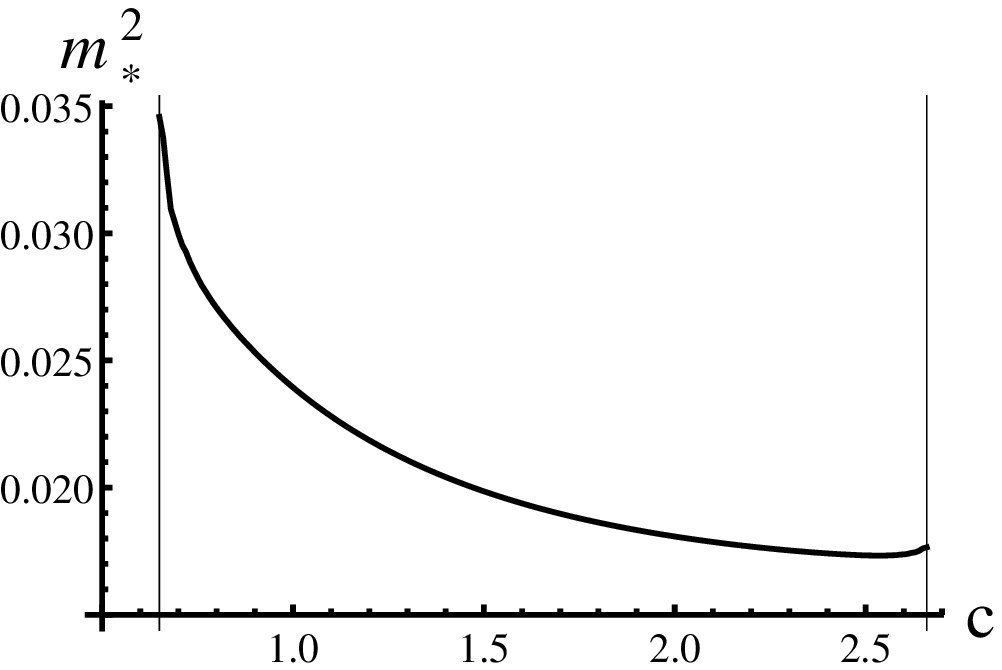}
\caption{\label{fig.4} Value of the dimensionless cosmological constant $\lambda_*$ (left), Newtons constant $g_*$ (center), and the 
massive graviton mass $m_*^2$ (right), arising from the UV-expansion of the fixed functions $\va_{*}(r)$ in the continuous window
\eqref{cwindow2}.}}

Substituting the ansatz \eqref{defans} into \eqref{pdgl} and expanding in $\epsilon$, 
the fixed function equation \eqref{fpeq} is recovered for $\epsilon = 0$. The information about relevant deformations
is encoded in the linear third order differential equation for $v(r)$ obtained at order $\epsilon$:
\be\label{defode}
\begin{split}
(3-\theta) v - 2 r v^\prime = & \, \frac{1}{\pi^2} \, \left( 1+ \tfrac{r}{6} \right)^{3/2} \, \frac{c_1 v^\prime + c_2 v^{\prime\prime} - \theta c_3 v^\prime - c_4 \left( \theta v^{\prime\prime} + 2 r v^{\prime\prime\prime} \right) }{3 \, \va_* + 4 \, (1-r) \, \va_*^\prime + 4 \, (2-r)^2 \, \va_*^{\prime\prime}} \\[1.2ex]
& \quad - \frac{\left(3 \va_* - 2 r \va_*^\prime \right) \left( 3 v + 4 (1-r) v^\prime + 4(2-r)^2 v^{\prime\prime} \right) }{3 \, \va_* + 4 \, (1-r) \, \va_*^\prime + 4 \, (2-r)^2 \, \va_*^{\prime\prime}} \, . 
\end{split}
\ee
Here the coefficients $c_i$ are given in \eqref{ccoeff} and we used that $\va_*$ satisfies the fixed function equation \eqref{fpeq} in order to simplify the resulting formula. Eq.\ \eqref{defode} depends explicitly on the fixed function $\va_*$ and parametrically on the stability coefficient $\theta$. Analogously to the study of fixed functions, relevant deformations $v(r)$ are given by solutions of \eqref{defode} which are regular on the entire interval $r \in [0, \infty[$. We expect that this regularity condition puts a constraint on the admissible values $\theta$, so that there is only a finite number of relevant deformations.

When searching for the values $\theta$ which give rise to regular solutions we proceed in complete analogy to Section \ref{sect:3.3}. Following the strategy
for the scalar case outlined in \cite{Morris:1994ie}, 
we first study the asymptotic behavior of \eqref{defode} for large $r$. In this limit, we identify 
the three fundamental solutions of the linear differential equation, which scale as
\be
v(r) \sim r^0 \, , \qquad v(r) \sim r^{113/92} \, , \quad \mbox{and} \qquad v(r) \sim r^{(3 - \theta)/2} \, , 
\ee 
respectively. The last solution is the asymptotic behavior expected from scaling arguments \cite{Morris:1994ie}. 
We thus fix the asymptotic behavior of the deformations according to $v(r) \sim r^{(3 - \theta)/2}$ and set
the coefficients for the other asymptotic solutions to zero.

After fixing the asymptotics of our solutions we integrate \eqref{defode} numerically and single out 
the parameters $\theta$ for which the solution remains regular at 
$x = x_{\rm sing}$. Restricting ourselves to the case of relevant deformations with $\theta > 0$ the
result is summarized in Table \ref{t.2}.
\TABLE[t]{
\begin{tabular}{||c||c|c|c|c||}
           & $c_{1, {\rm crit}}$ & $c_{2, {\rm crit}}$ & $c_{-, {\rm crit}}$ & $c_{+, {\rm crit}}$ \\ \hline \hline
\; \; $\theta_1$ \; \; & \; \; $4.896$  \; \; &  $ - $  & $ - $ & $ - $\\ \hline
\; \; $\theta_2$ \; \; & \; \; $0.287$  \; \; & \; $0.058$ \;  & \; $0.006$ \; & \; $0.002$ \;  \\ \hline
\end{tabular}
\caption{\label{t.2} The positive critical exponents associated with the UV-relevant deformations of 
the fixed functions shown in Fig.\ \ref{fig.3}.}
}
We find that the fixed function $\va_{1,*}(r)$ admits two relevant deformations while for the other fixed functions
there is only one admissible value $\theta > 0$. Thus the regularity condition indeed restricts the
allowed values of $\theta$ to a discrete and finite number. This central result constitutes strong evidence,
that the UV-critical surface of the corresponding fixed functionals is actually finite dimensional, despite
the inclusion of an infinite dimensional set of coupling constants in our truncation ansatz.

At this stage, the following remark is in order. The ansatz \eqref{defans} is restricted to
\emph{real} stability coefficients. In order to encode complex stability coefficients, eq.\ \eqref{defans} needs to 
be generalized to complex deformations $v(r) = v_{\rm Re}(r) + i v_{\rm Im}(r)$ according to
\be
\va(t, r) = \va_*(r) + \half \epsilon \, \left( e^{- ( \theta' + i \theta'') t} \, v(r) + {\rm c.c.} \right) \, . 
\ee
Substituting this generalized ansatz into \eqref{pdgl} results in a coupled system of third order differential equations
which depend on the real and imaginary part of $v(r)$ and \emph{two} parameters $\theta'$ and $\theta''$. We investigated
this system numerically, but were not able to find any regular deformations with complex critical exponents $\theta' + i \theta''$.

\section{Conclusions and outlook}
\label{sect.5}
In this work we used the exact functional RG equation \cite{Reuter:1996cp} to derive a non-linear partial differential
equation (PDE) governing the RG-flow of $f(R)$-gravity in conformally reduced Quantum Einstein Gravity (QEG) in
$d=3$ dimensions. In comparison to the previous works \cite{Machado:2007ea,Benedetti:2012dx},
our main technical improvement is the use of the exact heat-kernel formula on the sphere \cite{Avramidi:2000bm} which allows the construction 
of this PDE without exploiting specific properties of the regulator function $\cR_k$.

Our key interest is in scale-independent, regular solutions of this PDE which
in this particular case simplifies to an ordinary non-linear third order differential equation.
The resulting fixed functions constitute the generalization of the fixed points, known from
finite-dimensional truncations of the effective average action, to the realm
of infinitely many coupling constants. By combining analytic and numeric methods, we construct two isolated and one continuous one-parameter family of fixed functions. 
These constitute the first non-trivial fixed functions found in the gravitational framework. 
The effective action associated with the fixed functions has the form
\be\label{eq5.1}
\Gamma_*[g] = 6^{-3/2} \int d^3x \sqrt{g} \, c_{i, {\rm crit}} \, R^{3/2} \, ,
\ee
with the critical values $c_{i, {\rm crit}}$ given in \eqref{ciso2} and \eqref{cwindow2}. Notably,
this form of the effective action is the same as for the one-loop effective action found previously
\cite{Avramidi:2000bm}. 

For values of the RG-scale $k$ much larger than the curvature $R$, the solutions
admit a polynomial expansion of the form $f(R) = \sum_{i=0}^N \, \bar{u}_i \, R^i$. Based on
this expansion, all solutions give rise to a negative Newtons constant,
which is required for the positivity of the conformal factor kinetic term. Moreover, one of the discrete and all continuous fixed point solution have a positive massive graviton mass.
By studying linear deformations around the fixed function, we furthermore establish that they come with a finite number of UV-relevant deformations. Table \ref{t.2} indicates that there are two and one relevant deformations of these solutions respectively. Thus the UV-critical surface of the fixed functions is actually \emph{finite-dimensional}, despite the inclusion 
of an infinite number of coupling constants in the truncation. This result further strengthens the evidence that Asympotically Safe Quantum Gravity is actually a predictive
quantum theory of gravity with a small number of relevant coupling constants.

The physically most interesting properties are shown by the fixed function characterized by $c_{1, {\rm crit}}$. In addition to the properties found above, the corresponding fixed function is positive definite with a global minimum at zero curvature. Moreover, it can be shown that the polynomial expansion for RG-scales much larger than the curvature actually has a finite radius of convergence determined by the singularity structure of the underlying differential equation. 
Thus the polynomial $f(R)$-approximation, pioneered in \cite{Codello:2007bd} and subsequently refined in \cite{Machado:2007ea,Codello:2008vh,Bonanno:2010bt}, may give a valid and systematic approximatation of the underlying fixed function. We stress, that these properties are not put in ``by hand''. They arise dynamically when solving the fixed function equation.

At this stage, it is illustrative to compare our fixed functions
to previous studies of the gravitational RG-flow on finite-dimensional
subspaces. In \cite{Rechenberger:2012pm} the NGFP structure
arising from the beta-functions of three-dimensional QEG in the $R^2$-truncation
has been studied in detail. The corresponding flow equations 
took into account all metric fluctuations, but worked with a polynomial expansion up $R^2$ only.
The beta-functions on this three-dimensional truncation of the theory space give rise to one 
``physical'' NGFP and two ``unphysical'' NGFPs. Their properties are summarized in Table \ref{t.3}.

Comparing these results to the ones for conformally reduced QEG reported in Tables \ref{tab.4} and \ref{t.2}, we 
observe systematic differences: the conformal approximation and the full computation
systematically lead to different signs for the fixed point values for Newtons constant $g_*$ 
and the massive graviton mass $m_*^2$. In addition there is no clear relation between the stability coefficients $\theta$
found in the two computations. This lends itself to the interpretation
that conformally reduced gravity (as implemented in this paper) and the full theory may
actually probe different conformal field theories. 

In a sense, the results reported in Tables \ref{tab.4} and \ref{t.2}, are much closer to the properties of the matter-induced non-Gaussian fixed 
points studied in \cite{Codello:2008vh,Manrique:2010am,Codello:2011js}. Integrating out the quantum fluctuations
of a minimally coupled scalar field, the NGFP induced in the gravitational sector characteristically also appears
at negative Newtons constant $g_* < 0$. This is very much in the spirit of the proposal by 't Hooft \cite{Hooft:2010nc} that, 
from a path-integral perspective, the quantum fluctuations in the conformal factor should
be treated as matter fields. Their effect on the running of the gravitational coupling constants 
is qualitatively different from the contribution of the gravitational degrees of freedom defining the light-cone-structure of space-time. 
We hope to further clarify this point in the future.
\TABLE[t]{
\begin{tabular}{|c|ccc|ccc|}
          &  $\lambda_*$ & $g_*$ & $m^2_*$ & $\theta_1$ & $\theta_2$ & $\theta_3$ \\ \hline \hline
NGFP      & $0.019$ & $0.188$ & $-1.67$ & $8.39$ & $1.86$ & $1.35$ \\
UNGFP$_1$ & $0.364$ & $0.147$ & $-0.31$ & \multicolumn{2}{c}{$1.56 \pm 4.84 i$} & $-9.85$ \\
UNGFP$_2$ & $0.099$ & $0.216$ & $-0.22$ & \multicolumn{2}{c}{$0.19 \pm 0.97 i$} & $1.68$ \\ \hline
\end{tabular}
\caption{\label{t.3} Properties of the non-Gaussian fixed points found in the $R^2$-truncation of three-dimensional QEG including all metric fluctuations \cite{Rechenberger:2012pm}.}
}

We close our discussion with a remark on new massive gravity \cite{Bergshoeff:2009hq,Bergshoeff:2009aq}.
Comparing the expansion of our solutions for $r = R/k^2 \gg 1$ and $r \ll 1$, we note that their structure is manifestly different. 
For $r \ll 1$ our fixed functionals indeed have the structure of the new massive gravity Lagrangian, while
the exact form of the effective action is given in \eqref{eq5.1}. We expect that this actually constitutes an interesting observation
when discussing unitarity questions in this framework.

Clearly, it is desirable to generalize our results in various ways. Concerning the three-dimensional framework, the next logical step is the inclusion of the tensorial fluctuations. This would require generalizing the exact asymptotic heat-kernel formula \eqref{eq:heat-kernel} to higher order spin fields. This setup would lend itself to a direct comparison to the $N=2$ polynomial fixed point structure  \cite{Rechenberger:2012pm} summarized in Table \ref{t.3}. Of course, it would also be interesting to revisit the flow of $f(R)$-gravity in four dimensions \cite{Machado:2007ea,Benedetti:2012dx} where, so far, no complete solutions of the fixed function equation have been found. Based on the analysis of the RG-flow in the $R^2$-truncation \cite{Rechenberger:2012pm}, we expect that a detailed understanding of the zero-mode contributions originating from the transverse-traceless decomposition of the metric fluctuations will constitute a crucial prerequisite. 
 We hope to come back to these points in future work.

\acknowledgments
We thank D.\ Benedetti, A.\ Nink, R.\ Percacci and M.\ Reuter for helpful discussions. The research of F.~S.\ and O.~Z.\ is supported by the Deutsche Forschungsgemeinschaft (DFG)
within the Emmy-Noether program (Grant SA/1975 1-1).



\begin{thebibliography}{99}

\bibitem{Weinberg:1980gg} 
  S.~Weinberg
  in \textit{General Relativity, an Einstein Centenary Survey},
  S.W.~Hawking and W.~Israel (Eds.),
  Cambridge University Press, 1979; \\
  S.~Weinberg,
  hep-th/9702027.

\bibitem{Weinproc1}
  S.~Weinberg, arXiv:0903.0568; PoS C {D09} (2009) 001, arXiv:0908.1964.

\bibitem{Niedermaier:2006wt}
  M.~Niedermaier and M.~Reuter,
  Living Rev.\ Rel.\  {\bf 9} (2006) 5.

\bibitem{Reuter:2007rv}
  M.~Reuter and F.~Saueressig, in {\it Geometric and Topological Methods for Quantum Field Theory}, H.~Ocampo, S.~Paycha and
  A.~Vargas (Eds.), Cambridge Univ.\ Press, Cambridge, 2010, arXiv:0708.1317.

\bibitem{robrev}
  R.~Percacci, in \textit{Approaches to Quantum Gravity: Towards a New Understanding of Space, Time and Matter}, D. Oriti (Ed.), Cambridge University Press, Cambridge, 2009,
  arXiv:0709.3851.

\bibitem{Litim:2008tt}
  D.~F.~Litim, PoS(QG-Ph) {\bf 024} (2008), 
  arXiv:0810.3675.

\bibitem{Reuter:2012id}
  M.~Reuter and F.~Saueressig,
   New J.\ Phys.\ {\bf 14} (2012) 055022,
  arXiv:1202.2274.

\bibitem{Wetterich:1992yh}
  C.~Wetterich,
  Phys.\ Lett.\ B {\bf 301} (1993) 90.

\bibitem{Reuter:1996cp}
  M.~Reuter,
  Phys.\ Rev.\ D {\bf 57} (1998) 971,
  hep-th/9605030.

\bibitem{Bagnuls:2000ae}
  C.~Bagnuls and C.~Bervillier,
  Phys.\ Rept.\  {\bf 348} (2001) 91,
  hep-th/0002034.

\bibitem{Berges:2000ew}
  J.~Berges, N.~Tetradis and C.~Wetterich,
  Phys.\ Rept.\  {\bf 363} (2002) 223,
  hep-ph/0005122.

\bibitem{Delamotte:2007pf}
  B.~Delamotte,
  cond-mat/0702365.

\bibitem{Rosten:2010vm}
  O.~J.~Rosten,
  Phys.\ Rept.\  {\bf 511} (2012) 177,
  arXiv:1003.1366.
  
\bibitem{Pawlowski:2005xe}
  J.~M.~Pawlowski,
  Annals Phys.\  {\bf 322} (2007) 2831,
  hep-th/0512261.

\bibitem{Gies:2006wv}
  H.~Gies,
  hep-ph/0611146.

\bibitem{Braun:2011pp}
  J.~Braun,
  J.\ Phys.\ G {\bf 39} (2012) 033001,
  arXiv:1108.4449.

\bibitem{Dou:1997fg}
  D.~Dou and R.~Percacci,
  Class.\ Quant.\ Grav.\  {\bf 15} (1998) 3449,
  hep-th/9707239.

\bibitem{Souma:1999at}
  W.~Souma,
  Prog.\ Theor.\ Phys.\  {\bf 102} (1999) 181,
  hep-th/9907027.

\bibitem{Lauscher:2001ya}
  O.~Lauscher and M.~Reuter,
  Phys.\ Rev.\ D {\bf 65} (2002) 025013,
  hep-th/0108040.

\bibitem{Reuter:2001ag}
  M.~Reuter and F.~Saueressig,
  Phys.\ Rev.\ D {\bf 65} (2002) 065016,
  hep-th/0110054.

\bibitem{Litim:2003vp}
  D.~F.~Litim,
  Phys.\ Rev.\ Lett.\  {\bf 92} (2004) 201301,
  hep-th/0312114.

\bibitem{Lauscher:2001rz}
  O.~Lauscher and M.~Reuter,
  Class.\ Quant.\ Grav.\  {\bf 19} (2002) 483,
  hep-th/0110021.

\bibitem{Lauscher:2002sq}
  O.~Lauscher and M.~Reuter,
  Phys.\ Rev.\ D {\bf 66} (2002) 025026,
  hep-th/0205062.

\bibitem{Rechenberger:2012pm}
  S.~Rechenberger and F.~Saueressig,
  Phys.\ Rev.\ D {\bf 86} (2012) 024018,
  arXiv:1206.0657.

\bibitem{Codello:2007bd}
  A.~Codello, R.~Percacci and C.~Rahmede,
  Int.\ J.\ Mod.\ Phys.\ A {\bf 23} (2008) 143,
  arXiv:0705.1769.

\bibitem{Machado:2007ea}
  P.~F.~Machado and F.~Saueressig,
  Phys.\ Rev.\ D {\bf 77} (2008) 124045,
  arXiv:0712.0445.

\bibitem{Codello:2008vh}
  A.~Codello, R.~Percacci and C.~Rahmede,
  Annals Phys.\  {\bf 324} (2009) 414,
  arXiv:0805.2909.
  
\bibitem{Bonanno:2010bt}
  A.~Bonanno, A.~Contillo and R.~Percacci,
  Class.\ Quant.\ Grav.\  {\bf 28} (2011) 145026,
  arXiv:1006.0192.

\bibitem{Benedetti:2009rx} 
  D.~Benedetti, P.~F.~Machado and F.~Saueressig,
  Mod.\ Phys.\ Lett.\ A {\bf 24} (2009) 2233,
  arXiv:0901.2984;
  Nucl.\ Phys.\ B {\bf 824} (2010) 168,
  arXiv:0902.4630.

\bibitem{Reuter:2002kd}
  M.~Reuter and F.~Saueressig,
  Phys.\ Rev.\ D {\bf 66} (2002) 125001,
  hep-th/0206145.
  
\bibitem{Codello:2010mj}
  A.~Codello,
  Annals Phys.\  {\bf 325} (2010) 1727,
  arXiv:1004.2171.

\bibitem{Satz:2010uu}
  A.~Satz, A.~Codello and F.~D.~Mazzitelli,
  Phys.\ Rev.\ D {\bf 82} (2010) 084011,
  arXiv:1006.3808.

\bibitem{Eichhorn:2009ah} 
  A.~Eichhorn, H.~Gies and M.~M.~Scherer,
  Phys.\ Rev.\ D {\bf 80} (2009) 104003,
  arXiv:0907.1828;
  K.~Groh and F.~Saueressig,
  J.\ Phys.\ A A {\bf 43} (2010) 365403,
  arXiv:1001.5032;
  A.~Eichhorn and H.~Gies,
  Phys.\ Rev.\ D {\bf 81} (2010) 104010,
  arXiv:1001.5033.

\bibitem{Manrique:2010am}
  E.~Manrique, M.~Reuter and F.~Saueressig,
  Annals Phys.\  {\bf 326} (2011) 440, arXiv:1003.5129;
  Annals Phys.\  {\bf 326} (2011) 463, arXiv:1006.0099.

\bibitem{Becker:2012js} 
  D.~Becker and M.~Reuter,
  arXiv:1205.3583.

\bibitem{Benedetti:2010nr} 
  D.~Benedetti, K.~Groh, P.~F.~Machado and F.~Saueressig,
  JHEP {\bf 06} (2011) 079,
  arXiv:1012.3081.
  
\bibitem{Saueressig:2011vn}
  F.~Saueressig, K.~Groh, S.~Rechenberger and O.~Zanusso,
  PoS {\bf EPS-HEP2011} (2011) 124,
  arXiv:1111.1743.

\bibitem{Nink:2012vd}
  A.~Nink and M.~Reuter,
  arXiv:1208.0031.

\bibitem{Donkin:2012ud} 
  I.~Donkin and J.~M.~Pawlowski,
  arXiv:1203.4207.

\bibitem{Nagy:2012rn} 
  S.~Nagy, J.~Krizsan and K.~Sailer,
  arXiv:1203.6564.

\bibitem{Litim:2012vz}
  D.~Litim and A.~Satz,
  arXiv:1205.4218.

\bibitem{Morris:1994ie}
  T.~R.~Morris,
  Phys.\ Lett.\ B {\bf 329} (1994) 241,
  hep-ph/9403340.

\bibitem{Morris:1998da}
  T.~R.~Morris,
  Prog.\ Theor.\ Phys.\ Suppl.\  {\bf 131} (1998) 395,
  hep-th/9802039.

\bibitem{Canet:2003qd}
  L.~Canet, B.~Delamotte, D.~Mouhanna and J.~Vidal,
  Phys.\ Rev.\ B {\bf 68} (2003) 064421,
  hep-th/0302227.
  
\bibitem{Litim:2010tt}
  D.~F.~Litim and D.~Zappala,
  Phys.\ Rev.\ D {\bf 83} (2011) 085009,
  arXiv:1009.1948.

\bibitem{Flore:2012ma}
  R.~Flore, A.~Wipf and O.~Zanusso,
  arXiv:1207.4499.

\bibitem{Codello:2012sc} 
  A.~Codello,
  arXiv:1204.3877.

\bibitem{Cognola:2005de}
  G.~Cognola, E.~Elizalde, S.~Nojiri, S.~D.~Odintsov and S.~Zerbini,
  JCAP {\bf 02} (2005) 010,
  hep-th/0501096.

\bibitem{Benedetti:2012dx}
  D.~Benedetti and F.~Caravelli,
  JHEP {\bf 06} (2012) 017,
  arXiv:1204.3541.

\bibitem{Reuter:2008wj}
  M.~Reuter and H.~Weyer,
  Phys.\ Rev.\ D {\bf 79} (2009) 105005,
  arXiv:0801.3287.
  
\bibitem{Reuter:2008qx}
  M.~Reuter and H.~Weyer,
  Phys.\ Rev.\ D {\bf 80} (2009) 025001,
  arXiv:0804.1475.
  
\bibitem{Reuter:2009kq}
  M.~Reuter and H.~Weyer,
  Gen.\ Rel.\ Grav.\  {\bf 41} (2009) 983,
  arXiv:0903.2971.

\bibitem{Machado:2009ph}
  P.~F.~Machado and R.~Percacci,
  Phys.\ Rev.\ D {\bf 80} (2009) 024020,
  arXiv:0904.2510.

\bibitem{Bonanno:2012dg}
  A.~Bonanno and F.~Guarnieri,
  arXiv:1206.6531.

\bibitem{Avramidi:2000bm}
  I.~G.~Avramidi,
  Lect.\ Notes Phys.\ M {\bf 64} (2000) 1.

\bibitem{Hooft:2010nc}
  G.~'t Hooft,
  arXiv:1011.0061.

\bibitem{Groh:2011dw}
  K.~Groh, F.~Saueressig and O.~Zanusso,
  arXiv:1112.4856.

\bibitem{Codello:2012kq}
  A.~Codello and O.~Zanusso,
  arXiv:1203.2034.

\bibitem{Vilkovisky:1992za}
  G.~A.~Vilkovisky,
  {\it Heat kernel: Rencontre entre physiciens et mathematiciens},
  CERN-TH-6392-92.

\bibitem{Dowker:1975tf}
  J.~S.~Dowker and R.~Critchley,
  Phys.\ Rev.\ D {\bf 13} (1976) 3224.

\bibitem{Litim:2001up}
  D.~F.~Litim,
  Phys.\ Rev.\ D {\bf 64} (2001) 105007,
  hep-th/0103195.

\bibitem{Rahmede:2011zz}
  C.~Rahmede,
  PoS CLAQG {\bf 08} (2011) 011.

\bibitem{Narain:2009gb}
  G.~Narain and C.~Rahmede,
  Class.\ Quant.\ Grav.\  {\bf 27} (2010) 075002,
  arXiv:0911.0394.

\bibitem{Bergshoeff:2009hq}
  E.~A.~Bergshoeff, O.~Hohm and P.~K.~Townsend,
  Phys.\ Rev.\ Lett.\  {\bf 102} (2009) 201301,
  arXiv:0901.1766.
  
\bibitem{Bergshoeff:2009aq}
  E.~A.~Bergshoeff, O.~Hohm and P.~K.~Townsend,
  Phys.\ Rev.\ D {\bf 79} (2009) 124042,
  arXiv:0905.1259.

\bibitem{Ohta:2012vb}
  N.~Ohta,
  arXiv:1205.0476.

\bibitem{Codello:2011js}
  A.~Codello,
  New J.\ Phys.\  {\bf 14} (2012) 015009,
  arXiv:1108.1908.

\end{thebibliography}
\end{document}